\documentclass[aps,pra,twocolumn,floatfix,groupedaddress,superscriptaddress,nofootinbib,notitlepage]{revtex4-2}
\usepackage{times,bbm,bbold,amssymb,amsmath,amsfonts,dsfont,cancel,graphics,graphicx,color}
\usepackage[normalem]{ulem}
\usepackage{pifont}
\usepackage[usenames,dvipsnames]{xcolor}
\usepackage[colorlinks=true,citecolor=blue,linkcolor=blue,urlcolor=blue]{hyperref}

\DeclareFontFamily{OT1}{pzc}{}
\DeclareFontShape{OT1}{pzc}{m}{it}
              {<-> s * [1.25] pzcmi7t}{}
\DeclareMathAlphabet{\mathpzc}{OT1}{pzc}
                                 {m}{it}

\usepackage[T1]{fontenc} 
\usepackage{newtxtext,newtxmath} 

\def \llangle{{\langle\hskip-1mm\langle}}
\def \rrangle{{\rangle\hskip-1mm\rangle}}

\def\DSGamma{{\mathrm{I}\hskip-0.6mm\Gamma}}

\newcommand{\ignore}[1]{}
\newcommand{\ket}[1]{| #1 \rangle}
\usepackage[symbol*]{footmisc}

\DefineFNsymbolsTM{otherfnsymbols}{
\textasteriskcentered *
\textasteriskcentered *
}
\setfnsymbol{otherfnsymbols}

\begin{document}

\title{State-Based Quantum Simulation: Releasing the Powers of Quantum States and Copies}

\author{S. Alipour}
\email{s.alipoor@gmail.com}
\affiliation{Quantum Technology Finland Center of Excellence, Department of Applied Physics, Aalto University, P. O. Box 15600, FI-00076 Aalto, Espoo, Finland}

\author{A. T. Rezakhani}
\affiliation{Department of Physics, Sharif University of Technology, Tehran 14588, Iran}

\author{Alireza Tavanfar}
\affiliation{Champalimaud Research, Champalimaud Center for the Unknown, 1400-038 Lisbon, Portugal} 
\affiliation{Institute of Neuroscience, University of Oregon, Eugene, OR 97405, United States} 
\affiliation{Centre of eXplainable Artificial Intelligence, Technological University Dublin, Dublin, Ireland}

\author{K. M{\o}lmer}
\email{klaus.molmer@nbi.ku.dk}
\affiliation{Niels Bohr Institute, University of Copenhagen, Blegdamsvej 17, DK-2100 Copenhagen, Denmark} 

\author{T. Ala-Nissila}
\affiliation{Quantum Technology Finland Center of Excellence, Department of Applied Physics, Aalto University, P. O. Box 15600, FI-00076 Aalto, Espoo, Finland}
\affiliation{Interdisciplinary Centre for Mathematical Modelling and Department of Mathematical Sciences, Loughborough University, Loughborough, Leicestershire LE11 3TU, United Kingdom}

\begin{abstract}
Quantum computing employs controllable interactions to perform sequences of logical gates and entire algorithms on quantum registers. This paradigm has been widely explored, e.g., for simulating dynamics of manybody systems by decomposing their Hamiltonian evolution in a series of quantum gates. Here, we introduce a method for quantum simulation in which the Hamiltonian is decomposed in terms of states and the resulting evolution is realized by only controlled-swap gates and measurements applied on a set of auxiliary systems whose quantum states define the system dynamics. These auxiliary systems can be identically prepared in an arbitrary number of copies of known states at any intermediate time. This parametrization of the quantum simulation goes beyond traditional gate-based methods and permits simulation of, e.g., state-dependent (nonlinear) Hamiltonians and open quantum systems. We show how classical nonlinear and time-delayed ordinary differential equations can be simulated with the state-based method, and how a nonlinear variant of shortcut to adiabaticity permits adiabatic quantum computation, preparation of eigenstates, and solution of optimization tasks.
\end{abstract}
\date{\today}
\maketitle

\section{Introduction}
\label{sec:intr}

Quantum computers employ properties such as the superposition principle, nonlocality, and entanglement to calculate, process, or simulate quantum dynamics by a series of quantum \textit{gates} acting on parts of a composite quantum system. Despite the substantial utility of the gate-based quantum computation (GBQC) and simulation frameworks, a too rigid focus on the gate paradigm may divert our attention from less explored, yet available powers that quantum mechanics can bring to computation. 

Copying an \textit{unknown} quantum state is forbidden by the linearity of quantum theory. This no-cloning theorem rules out a wealth of over-optimistic proposals for applications of quantum effects, such as the use of entanglement for faster-than-light communication, and it severely restricts the value of quantum parallelism in superposition states in quantum computing, as well as the existence of simple schemes for quantum error correction. The no-cloning theorem, however, does not forbid the existence or preparation of any number of exact copies of a \textit{known} quantum state. Such copies or replica have well established applications in quantum physics \cite{Demler-1, Greiner-2, Bloch-1, Book:Nielsen-Chuang, QT, QPT-1,QPT-2, Huang-Preskill, Nakagawa, metro, Greiner-ent, Ekert-etal, measure-ent}. Along these lines, we note that the linear \textit{dynamics} of a quantum product state $\ket{\psi}\otimes \ket{\psi}$ is formally \textit{nonlinear} in the state amplitudes of the single-particle wave function $\ket{\psi}$. This nonlinearity is inherent in the quantum formalism of product states, and the use of numerous copies of an initial state permits \textit{state-based} quantum simulations that formally extend beyond the usual linearity restrictions of quantum simulators. This observation raises the following questions: Could states existing in several copies have applications in quantum technologies and lead to a paradigm shift from linear quantum computing to nonlinear simulation algorithms? 

In this article, we explore a new framework for quantum simulation and quantum computing which transcends traditional gate-based methods. In this \textit{state-based} quantum computing states play the primary role in the algorithm. Here gates and interactions on qubits are determined by the \textit{states} of auxiliary resource systems, provided when necessary during the algorithm. When any of the resource systems are prepared in the same state as the simulator system, the operations on the latter become formally and operationally nonlinear, with no deviation from the linearity of quantum mechanics or violation of the no-cloning theorem.  

In the following, we first introduce state-based quantum simulation (SBQS) for simulation of standard problems where the Hamiltonians are independent of the state of the system. We show that SBQS naturally applies to the simulation of nonunitary open-quantum system dynamics. We generalize the formalism and show that SBQS can also be used for simulation of Hamiltonians which are functions of the state of the system at current and/or past times. As an application, we show that SBQS with a state-dependent Hamiltonian permits solution of a set of \textit{classical} nonlinear delayed differential equations as exemplified by simulation of effective mean-field theories such as the Gross-Pitaevskii equation. Finally, we show that the effective nonlinearity in SBQS can significantly enhance adiabatic quantum computation and shortcut to adiabaticity and thereby be employed for simulation of various optimization problems.

\section{Quantum simulation paradigms}
\label{sec:QSim}

Quantum simulation \cite{Feynman} is a promising and rapidly developing field which uses quantum features to solve difficult problems in science and technology \cite{Lloyd:UQS, Aspuru, qsim-entropy, Nori, Preskill, QS-exp}. The most widely studied case is that of unitary evolution generated by a given Hamiltonian \cite{HHL, LinCombUnitaries, Poulin:TimeDepH, QHSBenchmark, QHS4NearTermHarware}. A large fraction of the proposed quantum simulation methods to date are gate-based, where a sequence of quantum gates are applied to simulate a given Hamiltonian. The sequence of gates in the GBQS method are chosen based on the so-called product formula \cite{QA4HS, ProductFormula} and can be used to study, e.g., quantum dynamics and to find ground or optimal states.  

However, there exist numerous important systems that are not easily amenable to gate-based methods, such as open quantum systems. Although quantum dynamics is linear, the evolution of an open-quantum system is generated by its correlations with its environment and such correlations may be represented by a formally nonlinear equation \cite{CorrPic}. Simulation of such systems is challenging for GBQS where a sequence of unitary quantum gates are applied. Moreover, the traditional quantum jump technique for simulation of open-system dynamics is basically applicable to linear evolution \cite{unraveling, unravelingNM} whereas it needs modifications and introduces a sample bias for nonlinear master equations \cite{Molmer-NME, Molmer-NME2}. Some other gate-based simulation methods have been recently proposed which are again basically designed for the linear case \cite{qChannel-4Unitary, LinearCombH}. Finally, there are numerous physical situations that lead to effective state-dependent Hamiltonians as natural approximations \cite{book:Kowalski, Gross-Pitaevskii, Kaplan} or to genuinely state-history-dependent (nonlinear) Hamiltonians which can emerge in complex manybody systems even at the level of unitary quantum dynamics \cite{ECQT-Tavanfar, ECQT}. 

There exist alternative methods for quantum simulation such as linear combinations of unitary evolutions, which are also based on application of quantum gates and are limited to linear systems \cite{LinCombUnitaries, UltraColdGases}. In addition, in most recent (quantum) machine-learning techniques for quantum simulation, the underlying quantum neural network is based on applying a sequence of parametrized quantum gates \cite{Molmer-learning, gate-based-qML1, gate-based-qML2, gate-based-qML3, Krenn}.  

Our state-based decomposition fundamentally changes the focus of the simulation from \textit{interactions} (and gates) to \textit{states}. In principle, this gives a significant advantage and opens vast possibilities in simulating various types of physical Hamiltonians without the need to explicitly simulate the interaction terms with quantum gate operations. This way the interactions are accounted for by coherences in the states which simulate the Hamiltonian.

By \textit{states} we mean density matrices or positive operators with unit traces in which the given auxiliary systems can be prepared. In addition to the auxiliary systems we also have the \textit{simulator} system on which the desired dynamics is going to be simulated. The states of the auxiliary systems are chosen such that the desired Hamiltonian can be decomposed in terms of their states. These quantum states can be chosen based on available 
resources in the lab for state preparation, due to which we call them \textit{resource states}. For example, we show that coherent quantum states or polarization states \cite{polarization-states} can be chosen as appropriate quantum states for the case of continuous and discrete quantum simulations, respectively. 
 
Another interesting feature of SBQS is that, in this scheme, we can use the power of multiple copies of the state of the system more naturally. In the SBQS method the instantaneous state of the evolving system, namely the simulator itself can also be part of the state-based algorithm. In other words, the current (or past) state(s) of the simulator system can also take part in its (time-dependent) Hamiltonian and hence its future evolution. This is the key feature that gives SBQS the ability to simulate state-dependent nonlinear evolution. 
 
\begin{figure}
\includegraphics[width=\linewidth]{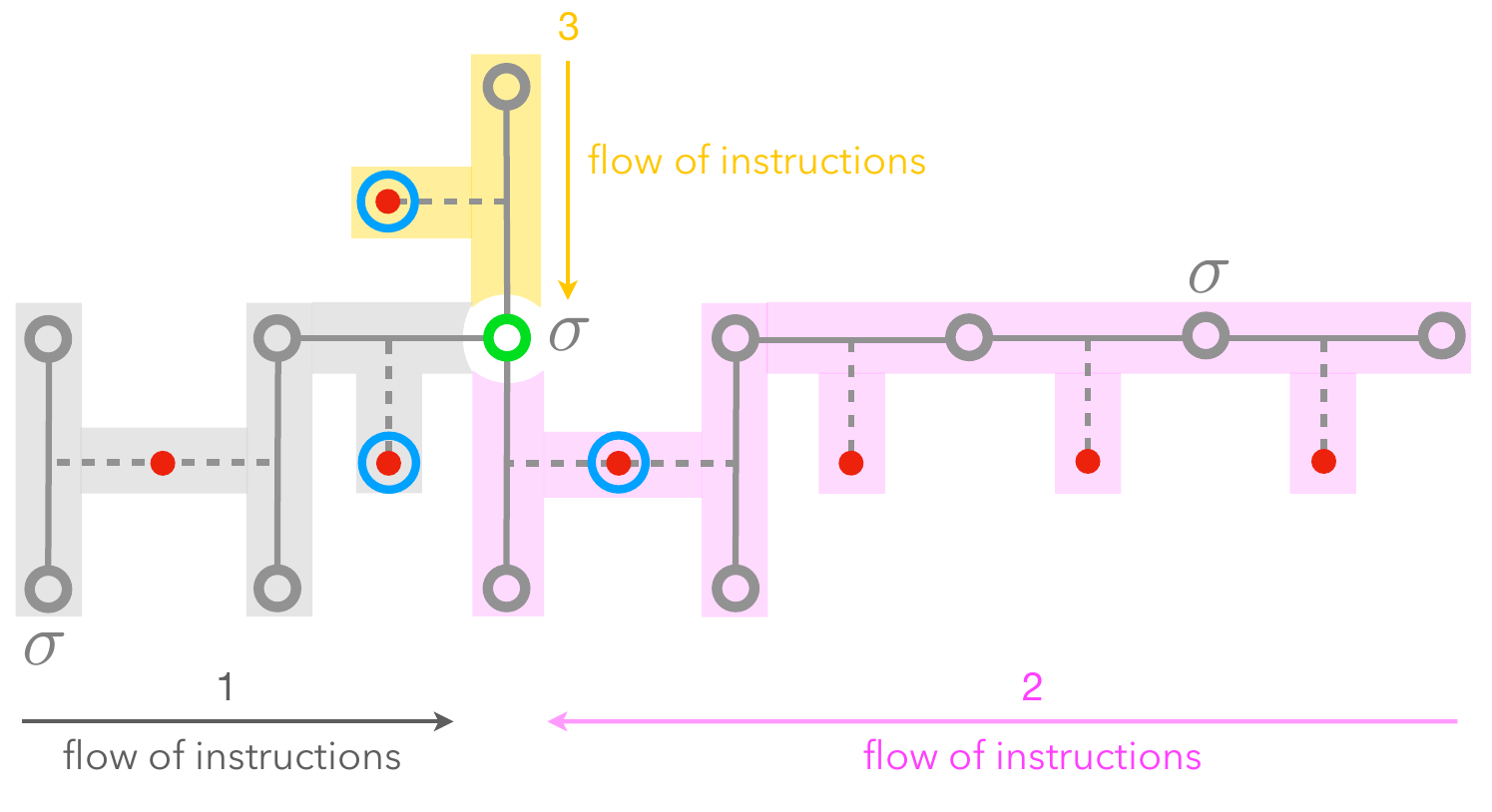}
\caption{Schematic of an SBQS algorithm. The green circle is the system of interest (simulator) whose dynamics is to be controlled and is initially in the state $\sigma$. All other systems with state $\sigma$ are copies of the system at the initial time. The other systems are ancillas and can initially be in an arbitrary state $\rho_{i}$. The red systems are used as control qubits. Each control qubit is connected with a dashed line to an edge (solid lines). The corresponding edge indicates the target systems on which the \textsc{swap} gate is applied. A blue circle around a state means a measurement is performed on that system. Here three sequential steps of a simulation are performed, each  highlighted by a different color and tagged with a time label such that step $n$ corresponds to simulation of the evolution for a given time interval $\delta t_{n}$. Preparation of several copies of the system in the initial state permits effective simulation of nonlinear evolution of the system.}
\label{fig:schematic}
\end{figure}

Using the density matrix exponentiation technique \cite{Lloyd-Mohseni-DME, DME-experimental} we show that by applying only controlled-\textsc{swap} (c-\textsc{swap} or Fredkin) gates \cite{cBS,cswap1,cswap2,cswap3} on the quantum states prepared combined with some post processing/post selection, it is possible to simulate both unitary and nonunitary evolution. We note that a common method for simulating nonlinear systems using GBQS is to first recast the problem as an equivalent \textit{linear} one (\textit{linearization}) \cite{Osborne-MultCop, linearization1, linearization2, Lloyd-Marvian-nonlinear-Diff, Childs-1, Childs-dissipative-NL, Krovi, Fujii, new-scaling, Umer-etal}. However, in SBQS we pursue a different approach. In Fig. \ref{fig:schematic} we illustrate a schematic of the SBQS design algorithm.

It is interesting to compare SBQS to the measurement-based quantum computation (MBQC) paradigm \cite{MBQC1, MBQC2}, which is a computationally equivalent alternative to GBQC \cite{MBQC-equiv-CBQC1, MBQC-equiv-CBQC2}. In MBQC a sufficiently entangled state is prepared and quantum algorithms are performed by applying measurements and single quantum gates on this state. SBQS has features from both GBQS and MBQC in that SBQS is state-based and it uses the the c-$\textsc{swap}$ gate, which is an entangling universal gate \cite{Book:Nielsen-Chuang}. SBQS leverages these two features to treat nonunitary open quantum system evolution as well as nonlinear (state-dependent) evolution. 

\section{SBQS for Hamiltonian simulation}
\label{sec:sbqs}

We first assume a canonical gate-based quantum computation scenario, where a given Hamiltonian $H(t)$ generates the unitary dynamics and is \textit{known} during the evolution. Later we will also show how SBQS can be used for more general cases, e.g., where $H$ can be a function of the instantaneous or the past states of the system and hence its form as a function of time is \textit{a priori} \textit{unknown}. We denote the states of the simulator with ``$\sigma$'' to distinguish it from the states ``$\rho$'' of the resource systems used for SBQS.
 
The SBQS method for a \textit{known} Hamiltonian simulation consists of the following three steps:\\

(\textbf{I}) \textit{Decomposition of $H$ in terms of quantum states}: A set of density matrices $\{\rho_{j}\}$ is chosen such that we can expand 
\begin{align}
H=\textstyle{\sum}_{j} h_{j} \rho_{j},
\label{H-StateDecomposition}
\end{align}
where $h_{j}$ are scalar coefficients and can be complex numbers in general. The set $\{\rho_{j}\}$ need not be complete or independent and can be chosen based on the ability to prepare particular quantum states. For technical convenience, we consider complete basis sets in most cases throughout this paper.\\

(\textbf{II}) \textit{Trotter-Suzuki expansion}. Up to this point, we have shown that any Hamiltonian can be decomposed in terms of quantum states, i.e., a form as in Eq. \eqref{H-StateDecomposition}. Next we define the operator $U=e^{-i t H}= e^{-i t \sum_{j} h_{j} \rho_{j}}$. We use the Trotter-Suzuki expansion to break down $U$ to a concatenation of factors with a single density matrix $\rho_{j}$ such that \cite{Lloyd:UQS}
\begin{align}
U = \big(\textstyle{\prod}_{j} e^{-i \delta h_{j} \rho_{j}}\big)^{n}+O(t^{2}/n),
\label{U-Trotter}
\end{align}
where $\delta= t/n$. This implies that to simulate $U$ it suffices to generate $e^{-i \delta h_{j} \rho_{j}}$.\\

(\textbf{III}) \textit{Exponentiation of density matrices}. To simulate the evolution $e^{-i \delta h \rho}$ on a quantum simulator system with the initial state $\sigma$, we apply the following steps \cite{Lloyd-Mohseni-DME}. First, another (auxiliary) quantum system with state $\rho$ is prepared. \\

Next a control qubit in the state $|\psi_{\delta}\rangle = |0\rangle -i \delta |1\rangle$ (with a sufficiently small $\delta$) is prepared and a c-\textsc{swap} gate $U_{\text{cs}}$ is applied \cite{Marvian}---which applies the \textsc{swap} gate $\mathcal{S}$ on the target systems when the qubit is in $|1\rangle$---yielding the state 

$U_{\text{cs}} \left(|\psi_{\delta} \rangle\langle \psi_{\delta}| \otimes (\rho \otimes \sigma) \right) U_{\text{cs}}^{\dag}$. After this operation the control qubit is measured with probability $p \approx 0.5$ to be in the state $|+_{01}\rangle = \frac{1}{\sqrt{2}}(|0\rangle + |1\rangle)$, and we obtain the heralded, renormalized  state with the reduced density matrix 
\begin{align}
\mathrm{Tr}_{1}[\langle +_{01}| & U_{\text{cs}} \left(|\psi_{\delta} \rangle\langle \psi_{\delta}| \otimes \rho \otimes \sigma \right) U_{\text{cs}}^{\dag} |+_{01}\rangle]/p \nonumber\\ 
&= \mathrm{Tr}_{1}[e^{-i \delta \mathcal{S}}(\rho\otimes \sigma) e^{i \delta \mathcal{S}} ] + O(\delta^{2})\nonumber\\
&= e^{- i \delta \rho} \sigma e^{i \delta \rho} + O(\delta^{2})
\label{eq-ID}
\end{align}
 
For other methods of bringing any Hamiltonian into the form \eqref{H-StateDecomposition}, see the Supplementary Information (SI).

As we argue later, this gives an advantage in simulating various interaction Hamiltonians without the need to explicitly consider the interaction terms with gate operations.

\textit{Example 1: Parametric amplifier}. The Hamiltonian of an optical parametric amplifier is given by $H=g a_{s}^{\dag}a_{i}^{\dag}+g^{*} a_{s} a_{i}$, where the subscripts ``$s$'' and ``$i$'' indicate signal and idler photons, respectively. Note that for simplicity we have omitted pump mode operators and assumed classical pumps \cite{NL-crystal, IndCohwithoutIndEm}. To simulate its dynamics we note that the Hamiltonian can be recast as $(1/\pi^{2}) \textstyle{\int_{\mathds{C}}} d^{2}\alpha \, d^{2}\beta\, \left(g\alpha^{*} \beta^{*}+ g^{*}\alpha\,\beta \right)|\alpha\rangle_{s}\langle \alpha| \otimes |\beta \rangle_{i}\langle \beta|$, which is in the desired form. To make it suitable for SBQS, we discretize the integrals so that the number of terms in the Hamiltonian becomes finite. See SI for details.
 
\section{SBQS of open quantum system dynamics}
\label{sec:sbqs-open}

Now we show how open quantum system dynamics can be simulated with the SBQS method. In the simplest approximation, the reduced dynamics of an open quantum system weakly coupled to a Markovian environment is given by a Lindblad dynamical equation $\dot{\sigma} = \mathpzc{L}[\sigma]$, where dot denotes $d/dt$ and the Lindbladian $\mathpzc{L}$ is given by 
\begin{align}
\mathpzc{L}[\sigma]=-i[H,\sigma] +\textstyle{\sum}_{k} \gamma_{k} \big(L_{k} \sigma L_{k}^{\dag} -(1/2) \{L_{k}^{\dag}L_{k}, \sigma\}\big).
\end{align}
To show that SBQS can also be applied on the open-system case we use vectorization, representing $A=\sum_{ij} a_{ij} |i\rangle\langle j|$ as $|A\rrangle:=\sum_{ij} a_{ij}|i\rangle \otimes| j\rangle$ in the computational basis $\{|i\rangle\}$. By vectorizing the Lindblad equation, an equation similar to the Schr\"odiner equation is obtained, which has a non-Hermitian generator, $\mathbbmss{L}$,  and acts on a larger Hilbert space. Following similar steps as for Hamiltonian simulation, the vectorized dynamics can be simulated by SBQS. For more details see Methods, and for an example see SI.  

\section{SBQS for designing more complex dynamical systems}
\label{sec:sbqs-dyn}

Thus far we have assumed that an arbitrary \textit{known} Hamiltonian is given, for which we delineated how to simulate the dynamics generated by that Hamiltonian employing the quantum states and e-\textsc{swap} (or c-\textsc{swap}) operations of SBQS. Now, we focus on a different problem: assuming that we can use \textit{arbitrary} quantum states $\{\rho_{j}\}$ and apply e-\textsc{swap} gates at will, what sort of dynamics can be simulated by using the SBQS method? We note that classical computers can also simulate what quantum systems or computers carry out, of course requiring typically significant overheads and resources. Now a similar question arises: how far can \textit{quantum} simulators go (with sufficient overheads and resources)? Here we show that SBQS indeed permit simulation of  dynamical systems beyond the realm of quantum mechanics. For example, we develop scenarios for simulating state-dependent (hence ``nonlinear'') Hamiltonian and open-system dynamics.

The key feature which endows SBQS with this potential is that in SBQS we can replicate the state of the system as one of the auxiliary states used in the decomposition of the Hamiltonian. This leads to the possibility of designing nonlinear quantum systems in contrast to the normal framework of quantum mechanics, where the evolution is linear and state-independent. Note that the apparent possibility of nonlinear dynamics to violate the no-cloning theorem is not an issue, as we explicitly provide the clones in the auxiliary system.  

\subsection{State-history-dependent (nonlinear) Hamiltonian simulation}
\label{sec:nonlinear}

We recall that nonlinear dynamical equations appear frequently in physics of (classical or quantum) manybody systems in various applications and contexts. For example, the Boltzmann equation \cite{book:Huang}, truncated Bogoliubov-Born-Green-Kirkwood-Yvon (BBGKY) master equations in atom optics and open quantum systems \cite{Meystre-NL, Molmer-NL, Los-NL, Rev-NL}, general open quantum system dynamics (correlation picture) \cite{CorrPic}, Hartree-Fock approximation \cite{Hartree-NL} and density functional theory (DFT) \cite{DFT-NL} in manybody physics and chemistry, Kardar-Parisi-Zhang (KPZ) equation in interface growth \cite{KPZ-NL}, and Navier-Stokes equations in hydrodynamics and weather forecasting \cite{NS-NL} are all nonlinear. Thus, developing techniques to solve nonlinear dynamical equations on quantum computers is of immediate scientific and practical significance. See, e.g., Refs. \cite{Osborne-MultCop, Lloyd-Marvian-nonlinear-Diff, Childs-1, Childs-dissipative-NL, Krovi, Fujii, new-scaling, Umer-etal} for previous attempts in this direction.

A natural extension of Eq. \eqref{H-StateDecomposition} is established by making the couplings depend on the states of the system rather than being constant parameters. This can happen when the Hamiltonian is state-history dependent, i.e., $H(t)=H(\{\sigma(t-a_{j})\}_j)$, where $\sigma(t-a_{j})$ is the state of the system of interest at (past or current) time $t-a_{j}$ with $a_{j}\geqslant 0$. It is evident that after decomposing this state-dependent Hamiltonian in terms of a chosen set of fixed density matrices $\{\rho_j\}$ the system-state dependence appears only in the coefficients of the decomposition, i.e., in the couplings $h_j$.

We consider the following specific form for this system-state dependence:
\begin{align}
h_{j}(\Gamma) = c_{j}\mathrm{Tr}[\xi_{j}\Gamma],
\label{hNL}
\end{align}
where $c_{j}$ is a constant real number, $\xi_{j}$ is a given constant trace-unity positive operator,  and we choose $\Gamma = \prod_{j} \sigma^{n_{j}}(t-a_{j})$ (constructed from the states of the system), with positive integer $n_{j}$'s and nonnegative $a_{j}$'s. This way we allow the state dependence of the Hamiltonian to be in the form of integer powers of the system state at some different times. We will show through several examples that this form covers a  large class of nonlinear problems. This significantly extends the scope of SBQS to state-history-dependent Hamiltonians. For a general formulation and investigation of this type of Hamiltonians---including a discussion of their emergence in natural and artificial systems---see Refs. \cite{ECQT-Tavanfar, ECQT}.

It is important to note that here the Hamiltonian is not \textit{known a priori} because the states of the system $\sigma(t-a_{j})$ are unknown before solving the dynamical equation. Only the prescription or description of the Hamiltonian in terms of the states of the system (at past or current times) $t-a_{j}$ is given. Thus, existing GBQS methods seem inadequate for simulation of such a dynamical system. To simulate this Hamiltonian using SBQS we first use the Trotter-Suzuki expansion, which means that we only need to simulate Hamiltonians of the form 
\begin{align}
H=c\,\mathrm{Tr}[\xi \Gamma] \,\rho.
\label{NLH}
\end{align}
 
\begin{figure}
\includegraphics[width=\linewidth]{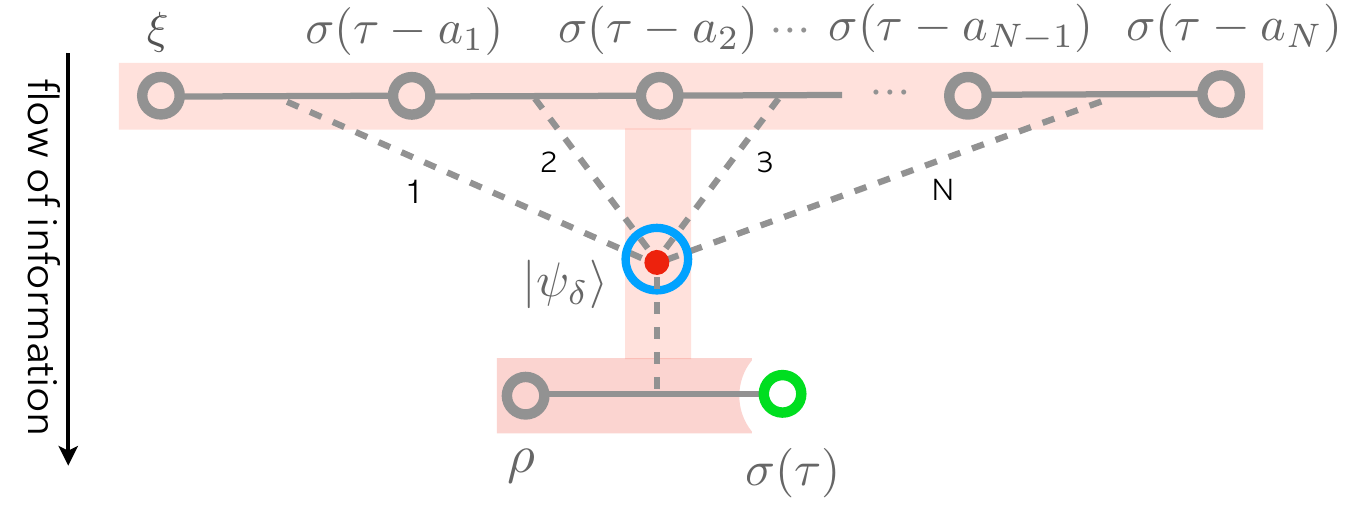}
\caption{The state-based quantum simulation of one time step of the dynamics governed by the Hamiltonian \eqref{NLH}: $N$ c-\textsc{swap} gates are applied respectively as shown in the figure on the corresponding pairs of target states, while the control qubit is common in all of them. The last c-\textsc{swap} gate is applied on the pair $\rho$ and $\sigma(\tau)$. Next, a projective measurement is applied on the control qubit after which all the systems except the simulator are traced out.}
\label{fig:NL-SBQS}
\end{figure}

Another important point is that the system evolving with this Hamiltonian has memory, i.e., its evolution at any time step depends on the state of the system at previous times. For solving the Schr\"{o}dinger equation of such systems---for simulating their history-dependent dynamics---it is not sufficient to have only the initial state of the system; instead, we need to know all the states in the initial time interval from $t=0$ to $t=\tau:=\max_{j} \{a_{j}\}$. With this information, we do the following steps for simulating the evolution of the system which is in the state $\sigma(\tau)$ to the next time step $\tau+\delta$, where $\delta \ll \min_{j}\{1,a_{j}\}$ (see Fig. \ref{fig:NL-SBQS}):\\
 
(\textbf{i}) State preparation: Prepare a control quantum system in the state 
$|\psi_{\delta}\rangle= |0\rangle -i c \delta |1\rangle$, 
which is normalized up to $O(\delta^{2})$ and is used as a control qubit for the application of a c-\textsc{swap} gate. Next, prepare the target systems in the state $\xi \otimes \DSGamma$, where
\begin{align}
\DSGamma = \otimes_{j} \sigma(\tau-a_{j})^{\otimes n_{j}}
\end{align}
is an $N$-partite system, with $N=\sum_{j} n_{j}$, and for each $j$ there are $n_{j}$ separate systems in the state $\sigma(\tau-a_{j})$. Hence we have $N+1$ target systems, which we label from $0$ to $N$. It should be noted that since the initial states from $t=0$ through the memory length $t=\tau$ are known and given, it is possible to make as many copies of these systems as we need.\\

(\textbf{ii}) Apply a concatenation of $N$ separate c-\textsc{swap} gates on adjacent target systems $\boldsymbol{U}_{\text{cs}}=U_{\text{cs}_{N-1,N}}\cdots U_{\text{cs}_{1,2}}U_{\text{cs}_{0,1}}$ (each \textsc{swap} gate is applied on the corresponding target systems when the control system is in the state $|1\rangle$) and then trace out over the target systems to obtain 
\begin{align}
\mathrm{Tr}_{\textsc{targets}}[\boldsymbol{U}_{\text{cs}}&(|\psi_{\delta}\rangle\langle \psi_{\delta}|\otimes  \xi \otimes \DSGamma) \boldsymbol{U}^{\dag}_{\text{cs}}] 
= |\tilde{\psi}_{\delta}\rangle \langle \tilde{\psi}_{\delta}| + O(\delta^{2}),
\label{post-selection}
\end{align}
where $|\tilde{\psi}_{\delta}\rangle=|0\rangle-ic\delta\, \mathrm{Tr}[\xi\Gamma] |1\rangle$ is the updated control system (see SI).\\ 

(\textbf{iii}) Apply another c-\textsc{swap} gate on the updated control system $|\tilde{\psi}_{\delta}\rangle$ and the new target systems in the state $\rho\otimes \sigma(\tau)$:
\begin{align}
U_{\text{cs}}& \big(|\tilde{\psi}_{\delta}\rangle\langle \tilde{\psi}_{\delta}|\otimes \rho\otimes \sigma(\tau)\big) U^{\dag}_{\text{cs}}=|0\rangle\langle 0|\otimes \rho\otimes \sigma(\tau) \nonumber\\
&-\Big(ic\delta\, \mathrm{Tr}[\xi\Gamma] |1\rangle\langle 0|\otimes \mathcal{S}\big(\rho\otimes \sigma(\tau)\big)+ \mathrm{h.c.}\Big) + O(\delta^{2}).
\end{align}
\\

(\textbf{iv}) Apply a projective measurement on the control system in the $\{|+\rangle, |-\rangle\}$ basis and keep the state only when the result of the measurement is $|+\rangle$. Then trace out the first target systems, which leads to the following state for the system:
\begin{align}
\sigma(\tau)-ic\delta\, \mathrm{Tr}&[\xi\Gamma][\rho,\sigma(\tau)] + O(\delta^{2}) \nonumber\\
& \approx e^{-i H \delta} \sigma(\tau) e^{i H\delta}  \equiv \sigma(\tau+\delta).
\end{align}
Note that in this step for simplicity we assumed that $H$ is Hermitian ($c\,\mathrm{Tr}[\xi\Gamma]\in\mathrm{I\!R}$). In the case of non-Hermitian $H$, the final result becomes $e^{-i H \delta} \sigma(\tau) e^{i H^{\dag}\delta}$.\\

(\textbf{v}) To go further in time, at some stage we will need the previous simulated states because the dynamics generated by the Hamiltonian (\ref{NLH}) may depend on some past times. Hence when necessary, we should repeat the simulation up to the instant that a copy of the system at previous history times is generated, so that we can use this copy to simulate the evolution in later times. Alternatively, we could have prepared sufficient auxiliary systems from the beginning and run the steps in parallel so that whenever we required copies of states at some earlier times we have them available.\\

\textit{Example 2: Gross-Pitaevskii Hamiltonian}. Bose-Einstein condensate systems with the Gross-Pitaevskii Hamiltonian are examples of systems with effective nonlinear unitary evolutions that emerge due to application of the Hartree approximation, i.e., mean-field theory, where $\dot{\sigma}(t) = -i [H_{0}+\sum_{i} \Pi_{i} \sigma(t) \Pi_{i} ,\sigma(t)]$ and $\Pi_{i}=|i\rangle\langle i|$ are projection operators in the real position space such that $\langle i| \sigma(t) |i\rangle$ is the local density at time $t$. To simulate this evolution using the Trotter-Suzuki expansion, we consider simulation of $H_{0}$ (which is a standard state-independent Hamiltonian) and $\sum_{i} \Pi_{i} \sigma(t) \Pi_{i}$ (which is a state-dependent Hamiltonian) separately. To simulate $H_{0}$, we can decompose it in terms of fixed quantum states and follow the steps in the previous section. To simulate $\sum_{i} \Pi_{i} \sigma(t) \Pi_{i}$, we note that it is equivalent to $\sum_{i} \mathrm{Tr}[\Pi_{i} \sigma(t)] \Pi_{i}$. Since $\Pi_{i} $ is already a quantum state, this Hamiltonian is exactly in the form of Eq. \eqref{NLH} and can be simulated with the instructions following this equation. For details see SI.

\subsection{Nonlinear delay-differential equations}

\textit{Nonlinear delay-differential equations} (NLDs) appear widely in physics, chemistry, engineering, biology, and social sciences, where systems feature delay or memory effects in their interactions \cite{ref:NLD}. In particular, recently quantum feedback control techniques and feedback-based universal approaches to photonic quantum computation have been proposed based on time-delay scenarios \cite{ref:NLD-2}. Nevertheless, due to intrinsic nonlinearities in dynamical systems with NLDs, it is not straightforward to simulate them by standard quantum simulation approaches. Here we develop a quantum algorithm based on SBQS to achieve this. 

Assume that a set of NLDs is given as
\begin{align}
\dot{x}_{m}(t) = \textstyle{\sum}_{n=1}^{D} \, f_{mn} \big(\boldsymbol{x}(t),\boldsymbol{x}(t-a_{1}),\cdots, \boldsymbol{x}(t-a_{N})\big)\, x_{n}(t),
\label{NLDDEq}
\end{align}
where dot denotes $d/dt$, $\boldsymbol{x}(t) = \big(x_{1}(t),x_{2}(t),\cdots,x_{D}(t)\big)^{T}$ is a vector of real variables at any time $t$, and $f_{mn}$ is a polynomial function of $D$ variables at $N$ different times, i.e., $f_{mn}=\mathrm{poly}(\ell_{0},\ell_{1},\ldots, \ell_{N})$, with $\ell_{j}$ being a natural number showing the highest degree of the variables at time $t-a_{j}$ and $a_{0}=0$. Any such $f_{mn}$ can be written generally as 
\begin{align}
f_{mn}=\textstyle{\sum}_{\{r_{ij}\}=\{1\}}^{\{\ell_{j}\}} C^{mn}_{\{r_{ij}\}} \textstyle{\prod_{j=0}^{N}} \prod_{i=1}^{D} \, x_{i}^{r_{ij}}(t-a_{j}), 
\label{f_{m}n-poly}
\end{align}
where at each memory time $t-a_{j}$, $\{r_{ij}\}$ is a set of $D$ integers indicating the degree of each variable $x_{i}$ at the given time. The first $D$ elements are the degrees of $D$ variables at time $t-a_{0}$, the second $D$ variables are the degrees of $D$ variables at time $t-a_{1}$, and so on. Each $r_{ij}$ is an integer between $0$ and $\ell_{j}$.

To show how SBQS can be used to solve Eq. \eqref{NLDDEq}, we map this problem to the simulation of a nonlinear dynamical system introduced in the previous section (to simplify the notation we drop time-dependence of variables where there is no risk of ambiguity). To do so, we first define a quantum state whose elements at time $t$ are constructed from the variables of the NLDs at the same time so that $
|\psi(t)\rangle=\textstyle{\sum_{i=0}^{D+1}} x_{i}(t)|i\rangle$, where $x_0(t)$ is constant and $x_{D+1}(t)$ is used for normalization of the state. By rewriting the NLDs in terms of $|\psi\rangle$ at different times, we obtain a Schr\"{o}dinger equation with a state-history-dependent Hamiltonian, which can be simulated using SBQS. For more details and another example (logistic map) see Methods. 

\subsection{Digital-analog adiabatic quantum computation}
\label{sec:aqc}

The unique feature of SBQS being able to accommodate state-dependent Hamiltonians enables us to significantly enhance adiabatic quantum computation (AQC) \cite{AQC} by generalizing ideas from ``shortcut to adiabaticity'' (STA) \cite{CD}. AQC is based on the adiabatic theorem \cite{exp-ad}, which requires a sufficiently slow and gapped time-dependent Hamiltonian. There the solution of a ``hard'' problem is given by the ground state of a hard Hamiltonian, assuming that we start from the ground state of a ``simple'' Hamiltonian.

In contrast, STA is a quantum control technique to ensure that the dynamics keeps us in a specific eigenstate (e.g., the ground state) of a given time-dependent Hamiltonian $H(t)$ in a finite time \cite{CD, Refael-etal}. There if the system is prepared in the ground state $G(0)$ of a Hamiltonian $H(0)$, the evolution generated with the ``counterdiabatic'' Hamiltonian $H_{\textsc{cd}}(t) = H(t) + i[\dot{G}(t), G(t)]$, where $G(t)$ is the projection on the instantaneous ground state of $H(t)$, keeps the system in $G(t)$ at all times, i.e., $\sigma(t)=G(t)$. Notwithstanding its utility, this algorithm presumes that $G(t)$ is already \textit{known} while it may, in contrast, be  \textit{a priori unknown}, and indeed an objective of AQC. In this sense, STA is not immediately applicable to AQC. Here we show how to overcome this obstacle and simulate the ground state of a Hamiltonian at arbitrary finite times. This unites and combines the forces of AQC and STA.

We generalize $H_{\textsc{cd}}(t)$ by replacing $G(t)$ with $\sigma(t)$,
\begin{align}
\widetilde{H}_{\textsc{cd}}(t) = H(t) + i[\dot{\sigma}(t), \sigma(t)].
\label{CD-Ham}
\end{align}
Although this Hamiltonian is different from the original $H_{\textsc{cd}}(t)$, we observe that this one also leads to the desired evolution if the initial state of the system is an eigenstate of the initial Hamiltonian $H(0)$. Specifically, we prove that if a \textit{pure} state satisfies the Schr\"{o}dinger equation with $\widetilde{H}_{\textsc{cd}}(t)$ ($\forall t\in[0,T]$), this state can only be an eigenstate of $H(t)$ (cf. SI). Using the approximation $\dot{\sigma}(t) \approx \big(\sigma(t)-\sigma(t-\tau)\big)/\tau$, for small $\tau$, recasts
\begin{align}
\widetilde{H}_{\textsc{cd}}(t,\sigma) \approx H(t) - (i/\tau)[\sigma(t-\tau), \sigma(t)],
\end{align}
which is now within the class of the Hamiltonians that can be simulated by SBQS as in Sec. \ref{sec:nonlinear}. This way we \textit{digitize} a counterdiabatic evolution for STA without needing the knowledge of the instantaneous ground state $G(t)$ of the Hamiltonian $H(t)$ \textit{a priori}. Obviously this approach can be equally used to find any eigenstate of an arbitrary observable and also it can be tailored for various optimization tasks.  

For example, this is the very objective of variational quantum eigensolver (VQE) simulations, which compute the ground state of a given observable, $\min_{\psi}\langle \psi|A|\psi\rangle$, by employing a mixture of quantum and classical computations \cite{NISQ-, VQE-rev, VQE-state, HF-Google}. The SBQS technique can be applied to this type of optimizations and even more general ones of the form $\min_{\psi}\langle\psi|A(\psi)|\psi\rangle$. 

The DFT problem \cite{DFT-NL} can also be recast in SBQS. DFT is the most powerful and principal computational toolbox of \textit{many-electron} problems in quantum chemistry and materials simulations \cite{DFT-hard, Hastings, Miessen, DFT-QC, Qmanybody}. This approach reduces the original manybody problem of finding the ground-state energy to an equivalent effective single-electron problem. In principle, in DFT one aims to find the ground state (energy) of a single-electron, \textit{state-dependent} Hamiltonian $H(\sigma) = V + F(\sigma)$, where $V$ is an external potential and $F(\sigma)$ is a universal functional which accounts for the kinetic energy and electron-electron interactions of the electronic manybody system. However, not only the exact form of $F(\sigma)$ is \textit{unknown}, but also obtaining it is computationally hard \cite{DFT-hard}. To alleviate this problem, in practice DFT resorts to approximations, which amounts to choosing some ansatz for the form of $F(\sigma)$. In such cases that the form of (approximate) $H(\sigma)$ becomes known, we can apply SBQS in the following counterdiabatic manner. We start with an easy Hamiltonian $H_{0}$ and its ground state $|\phi_{0}\rangle$ and consider the transient Hamiltonian $H \big(s(t),\sigma\big) = \big(1-s(t)\big) H_{0} + s(t) \, H(\sigma)$, where $s(0)=1-s(T)=0$ (for some smoothly controllable $s(t)$ with $T>0$). We then follow Eq. (\ref{CD-Ham}) to simulate the counterdiabatic \textit{state-dependent} Hamiltonian $\widetilde{H}_{\textsc{cd}}(t,\sigma) = H\big(s(t),\sigma\big) - (i/\tau)[\sigma(t-\tau), \sigma(t)]$. 

In SI we also show how to minimize the variance of an observable. To our knowledge, achieving such goals is not immediately clear within GBQS. 

\section{Summary and outlook}
\label{sec:outlook}

We have proposed a state-based quantum simulation (SBQS) technique which rather than decomposing Hamiltonians in terms of local interactions and using associated quantum gates hinges on decomposing Hamiltonians in terms of quantum states and using density matrix exponentiation. SBQS substantially enhances the potential of quantum simulation and transfers the complexity to the ability of an experimental setup to prepare customized quantum states rather than execute quantum gates.

Most importantly, we have shown that in SBQS it is possible to use the very states of the quantum simulator system itself as parts of the simulation algorithm. This way current and past states of the simulator can affect its future evolution. This key feature endows SBQS with the ability to simulate state-dependent, nonlinear evolutions. In particular, SBQS allows for solving nonlinear time-delayed classical and quantum problems without the need for the linearization. Noting that linearization (as a common tool for solving nonlinear equations in gate-based simulation) is computationally costly, this feature of SBQS can reduce simulation costs considerably. As a result, we have proven that SQBS can help combine adiabatic quantum computation and shortcut to adiabaticity in order to tackle various optimization problems such as variational quantum algorithms and DFT.

SBQS is a natural quantum simulation approach for a broader class of classical and complex quantum problems as compared to the widely used gate-based method. The natural characteristics of SBQS make it conducive to simulating and solving diverse problems and models in realistic complex systems. Indeed, complex, state-history-dependent, and nonlinear processes and phenomena are ubiquitous and significant, spanning weather forecast, aerodynamics, hydrodynamics, chemistry, pharmaceutics, open-system dynamics, and biological and cognitive phenomena. There exist vast scientific and technological problems which can immediately benefit from the versatility and utility of SBQS.

\textit{Acknowledgments.---}This work was partially supported by the Academy of Finland's Center of Excellence program QTF Project 312298 (to T.A.-N.) and by the Danish National Research Foundation Centre of Excellence ``Hy-Q,'' Grant No. DNRF139 (to K.M.). Some elements of this work are included in a patent filed by Aalto University.


\onecolumngrid

\clearpage
\begin{center}{\textbf{Methods}}
\end{center}

\section{Decomposition of $H$ in terms of quantum states}

(\textit{a}) \textit{Discrete case}: \textit{Decomposition of $H$ in terms of polarization states}. We start with the representation of $H$ in the computational basis (or number-state basis) $H=\sum_{mn} \langle m | H | n \rangle |m \rangle \langle n|$. Using the polarization identity for this basis \cite{Book:Nielsen-Chuang} 
\begin{align}
|m \rangle \langle n| = |\mathsf{+}_{mn}\rangle\langle\mathsf{+}_{mn}| +i |\mathsf{-}_{mn}\rangle\langle\mathsf{-}_{mn}| - \textstyle{\frac{1+i}{2}} (|m\rangle\langle m| + |n\rangle\langle n|),
\label{polarization_bases}
\end{align}
where 
$|\mathsf{+}_{mn}\rangle=(|m \rangle + |n \rangle)/\sqrt{2}$ and $|-_{mn}\rangle=(|m \rangle+i|n \rangle)/\sqrt{2}$, we can find a quantum-state decomposition for $H$ as  
\begin{equation}
H=\textstyle{\sum}_{mn} H_{mn} \big(|\mathsf{+}_{mn}\rangle\langle\mathsf{+}_{mn}| +i |\mathsf{-}_{mn}\rangle\langle\mathsf{-}_{mn}|-\frac{1+i}{2}(|m\rangle\langle m|+|n\rangle\langle n|)\big).
\label{H-polarization}
\end{equation}
\\
(\textit{b}) \textit{Continuous case}: \textit{Decomposition in terms of coherent states}. By using the Glauber-Sudarshan $P$ representation  theorem \cite{Book:Scully-Zubairy, Fan, Goldberg}, it is immediate that any Hermitian operator can be decomposed \textit{diagonally} in terms of (normalized) coherent quantum states $|\alpha \rangle \langle \alpha|$. Thus, for any Hermitian $H$ we have
\begin{align}
H = \textstyle{\int_{\mathds{C}}} d^{2}\alpha \, h(\alpha, \alpha^{*}) \, |\alpha \rangle\langle \alpha|,
\label{H-coherent}
\end{align}
where $h(\alpha, \alpha^{*}) = (e^{|\alpha|^{2}}/\pi^{2}) \textstyle{\int_{\mathds{C}}} d^{2} \beta \, \langle - \beta| H |\beta \rangle \, e^{|\beta|^{2} - \beta \alpha^{*} + \beta^{*} \alpha}$. We remark that although in general computing $h(\alpha, \alpha^{*})$ involves derivatives of highly singular functions such as the Dirac delta function, specific remedies to alleviate this difficulty have also been proposed \cite{Klauder, Lobino, Keshari-Rezakhani}.

\section{SBQS for open quantum systems}

(\textbf{i}) Assuming that a known Lindblad superoperator is given, we use a vectorization technique to transform it to an operator acting on a higher dimensional Hilbert space. An operator $A$ represented as $A=\sum_{ij} a_{ij} |i\rangle\langle j|$ in the computational basis $\{|i\rangle\}$ is transformed to a vector $|A\rrangle:=\sum_{ij} a_{ij}|i\rangle \otimes| j\rangle$. Using this definition and the identity $|A B C\rrangle= (A\otimes C^{T})|B\rrangle$ (with $T$ being transposition), we can obtain the vectorized representation of $\dot{\sigma} = \mathpzc{L}[\sigma]$ as  
\begin{align}
\label{Lindblad_vectorized}
|\dot{\sigma}\rrangle = \mathbbmss{L}|\sigma\rrangle,
\end{align}
where $|\sigma\rrangle$ is the vectorized form of the density matrix of the system and $\mathbbmss{L}$ is the counterpart of $\mathpzc{L}$ in the vectorized space (for details of vectorization see Ref. \cite{DQM}). We use this vectorized relation to implement SBQS.\\

(\textbf{ii}) We prepare a quantum system in the initial state $|\psi(0)\rangle=|\sigma(0)\rrangle / \sqrt{\llangle \sigma(0)|\sigma(0)\rrangle}$. We note that although $\sigma$ is a density matrix, its vectorization $|\sigma\rrangle$ is not necessarily a normalized quantum state vector. The normalization in $|\psi(0)\rangle$ should be considered as a scaling of the variables of Eq. \eqref{Lindblad_vectorized}, $\sigma_{ij} \to \sigma_{ij}/\sqrt{\llangle \sigma(0)|\sigma(0)\rrangle}$, while the equation remains unchanged with the scaled variables.\\

(\textbf{iii}) Equation \eqref{Lindblad_vectorized} is similar to the Schr\"{o}dinger equation with a non-Hermitian Hamiltonian. Similar to what we did previously for simulation of such dynamics, we first expand $\mathbbmss{L}$ in terms of quantum states 
\begin{equation}
\mathbbmss{L}=\textstyle{\sum}_{r} l_{r} \rho_{r},
\end{equation}
where the $\rho_{r}$ are quantum states in the bipartite vectorization space. This equation is similar to Eq. \eqref{H-StateDecomposition} for which we have already developed our simulation protocol. Naturally extending the steps of simulation of linear Hamiltonian dynamics, here we obtain 
$|\psi(0)\rangle \to e^{t \mathbbmss{L}} |\psi(0)\rangle =|\sigma(t)\rrangle/{\sqrt{\llangle \sigma(0)|\sigma(0)\rrangle}}$. 
However, the quantum state that describes the system in any measurement is instead the normalized quantum state 
\begin{align}
|\psi(t)\rangle=|\sigma(t)\rrangle/\sqrt{\llangle \sigma(t)|\sigma(t)\rrangle}.
\end{align}
\\

(\textbf{iv}) Next we should read the information of the \textit{physical} state $\sigma(t)$ from the \textit{simulated} state $|\psi\rangle$. To this end, we note that $\sum_{i}\langle i,i|\sigma\rrangle=\mathrm{Tr}[\sigma]=1$ and hence
\begin{align}
\sigma=\frac{\sum_{ij} \langle i, j|\sigma\rrangle |i\rangle\langle j|}{\sum_{k} \langle k, k|\sigma\rrangle}=\frac{\sum_{ij} \langle i,j|\psi\rangle |i\rangle\langle j|}{\sum_{k} \langle k, k |\psi\rangle}.
\end{align} 

To obtain the expectation value of an observable $A$ of the system of interest (with density matrix $\sigma$) from measurements on the simulator (populating the quantum state $|\psi \rangle$) we note that $A$ can be written as $A=\sum_{i}^{r} a_{i} P_{i}$, with eigenvalues $a_{i}$ and corresponding eigenprojectors $P_{i}$. The expectation value is $\langle A\rangle = \sum_i a_i p_i$, where $p_{i}=\mathrm{Tr}[P_{i} \sigma]=\llangle P_i| \sigma \rrangle$. We note that $p_{i} 
^2 =\mathrm{Tr}[P_i \sigma]^2=\llangle \sigma| \mathbbmss{P}_i| \sigma \rrangle \propto \langle \psi |\mathbbmss{P}_i|\psi \rangle$, where $\mathbbmss{P}_i=|P_i\rrangle \llangle P_i|$. Measuring the state of the simulator in the orthonormal basis $\{|P_i\rrangle\}$ thus  yields the different eigenstates with relative probabilities $\propto p_i^2$, and permits evaluation of the (normalized) probabilities $p_i$ and hence the mean value of $A$.    
\section{SBQS for nonlinear delay-differential equations}

We propose an SBQS algorithm to simulate NLDs. We need to follow the following steps:\\

(\textbf{i}) We add two extra equations to the set in Eq. \eqref{NLDDEq}. One equation is added for a new independent fixed parameter $x_{0}$: $\dot{x}_{0}=0$, which implies $f_{00}=1$ and $f_{m0}=f_{0n}=0,\,\forall m,n$. This constant parameter has been introduced to take care of some technicalities which will become clear later. Another equation is added for a normalizing variable $x_{D+1}=\textstyle{\sqrt{1-\sum_{i=0}^{D} x_{i}^{2}}}$, such that $\dot{x}_{D+1} = -x_{D+1}^{-1} \boldsymbol{x} \cdot \dot{\boldsymbol{x}} = -x_{D+1}^{-1} \sum_{m,n=1}^{D} f_{mn} x_{m} x_{n}=:f_{D+1,0}\,x_{0}$, where $f_{D+1,0} =-(x_{0} x_{D+1})^{-1} \sum_{m,n=1}^{D} f_{mn} x_{m} x_{n}$. Using the Taylor expansion about $x_{D+1}=1$ we obtain $x_{D+1}^{-1}= \sum_{k=0}^{r} (1-x_{D+1})^{k} + O\big((1-x_{D+1})^{r}\big)$, where $r$ is chosen according to our desired error. Thus,  
$f_{D+1,0}$ becomes a polynomial of all variables similar to the expression for $f_{mn}$ in Eq. \eqref{f_{m}n-poly}, but where index $i$ runs from $0$ to $D+1$. If necessary, we can first scale all variables so that normalization becomes possible. Now we define a normalized quantum state as
\begin{align}
\label{var_vec}
|\psi(t)\rangle=\big(x_{0}, x_{1}(t),\cdots,x_{D}(t), x_{D+1}(t)\big)^{T} = \textstyle{\sum_{i=0}^{D+1}} x_{i}(t)|i\rangle.
\end{align}
It is straightforward to see that the set of differential equations can be recast in the form of a Schr\"{o}dinger-like equation $|\dot{\psi}(t)\rangle=-i H_{\textsc{nld}} |\psi(t)\rangle$, where  
\begin{align}
H_{\textsc{nld}}=i\textstyle{\sum}_{m,n=0}^{D+1} f_{mn}\big(\boldsymbol{x}(t),\boldsymbol{x}(t-a_{1}),\cdots, \boldsymbol{x}(t-a_{N})\big) |m\rangle\langle n|
\label{H-nonlinearEq}
\end{align}
is a \textit{nonlinear} (variable-dependent) operator which is not necessarily Hermitian. Nonetheless, SBQS can successfully simulate its dynamics.\\ 

(\textbf{ii}) To transform $H_{\textsc{nld}}$ into the form of Eq. \eqref{H-StateDecomposition} with nonlinear couplings, it is sufficient to write $f_{mn}$ in the form of Eq. \eqref{hNL}. We note that if $r_{ij}$ is even, then $x_{i}^{r_{ij}}(t-a_{j})=\big(\langle \psi(t-a_{j})|i \rangle\langle i|\psi(t-a_{j}) \rangle\big)^{r_{ij}/2}$, which is the expectation value of the operator $|i\rangle\langle i|^{\otimes r_{ij}/2}$ with respect to multiple copies of the state of the system at relevant times, $|\psi(t-a_{j})\rangle^{\otimes r_{ij}/2}$. When $r_{ij}$ is odd, using the constant parameter $x_{0}$ we can also write $x_{i}^{r_{ij}}(t-a_{j})$ in terms of the expectation value of an operator: $x_{i}^{r_{ij}}(t-a_{j})= \big(\langle \psi(t-a_{j})|i \rangle\langle i|\psi(t-a_{j})\rangle\big)^{\lfloor {r_{ij}}/{2} \rfloor} \big(x_{0}^{-1}\langle \psi(t-a_{j})|0 \rangle\langle i|\psi(t-a_{j})\rangle\big)$, which is the expectation value of the non-Hermitian operator 
$x_{0}^{-1}|i\rangle\langle i|^{ \otimes\lfloor r_{ij}/2\rfloor}\otimes |0\rangle\langle i|$ with respect to the state $|\psi(t-a_{j}) \rangle^{\otimes\lfloor r_{ij}/2\rfloor+1}$. Here $\lfloor \cdot \rfloor$ denotes the floor function. For simplicity and without loss of generality, we assume in the following that all $r_{ij}$s are even. Hence we can rewrite any $f_{mn}$ given in Eq. \eqref{f_{m}n-poly} as the expectation value of the operator 
\begin{align}
F_{mn}=\textstyle{\sum}_{\{r_{ij}\}=\{1\}}^{\{\ell_{j}\}} C^{mn}_{\{r_{ij}\}} \textstyle{\otimes_{j=0}^{N} \otimes_{i=0}^{D+1} }  | i\rangle\langle i|^{\otimes r_{ij}/2},
\end{align}
with respect to the quantum state
\begin{align}
\label{Psi}
|\Psi\rangle:=\textstyle{\otimes}_{j=0}^{N} |\psi(t-a_{j}) \rangle^{\otimes  (D+2) \ell_{j}/2},
\end{align}
such that $f_{mn}=\langle\Psi|F_{mn}|\Psi \rangle$. Equivalently, using the \textsc{swap} identity we can rewrite 
\begin{align}
f_{mn}=\mathrm{Tr}[\mathcal{S}\,(F_{mn}\otimes|\Psi \rangle\langle\Psi|)].
\end{align}
Using this, it is evident that all terms in $H_{\textsc{nld}}$ can be written in the desired form of Eq. \eqref{NLH} and hence can be simulated using SBQS.  \\ 

As a final remark in this part, note that because in the steps of the simulation explained in Sec. \ref{sec:nonlinear} all traces are taken over different parties, we can alternatively defer all traces until the end of the protocol. In addition, SBQS bypasses the need for computation of intermediate averages by a suitable post-selection [Eq. (\ref{post-selection})]. These features may offer some advantages for practical implementations of the simulation---see, e.g., remarks made in Ref. \cite{Molmer-NME2}.

\textit{Example 3: Logistic map}. Consider the logistic map differential equation $\dot{x}= (r-1) x-r x^{2}$. Here we need three copies of the state of the system at each time to simulate every single term of the Trotter-Suzuki expansion, which includes several terms. It should be noted that the number of Trotter-Suzuki terms depends on the decomposition of the Hamiltonian in terms of quantum states. Here we have performed this with a brute-force approach and as a proof of principle in order to demonstrate the direct, explicit implementation of SBQS. Details have been relegated to SI.\\

\textbf{Data availability}. The data that support the findings of this study are available in the manuscript and SI.

\clearpage
\begin{center}
\textbf{Supplementary Information}
\end{center}
\setcounter{section}{0}
\section{Other approaches to Hamiltonian decomposition in terms of quantum states}
\label{app:others}

\subsection{Projecting a Hamiltonian onto its positive and negative supports}

It is always possible to write any given Hermitian operator $H$ as 
\begin{align}
H=H^{(+)}-H^{(-)},
\end{align}
where $H^{(+)}=P^{(+)} H P^{(+)}$ and $H^{(-)}=-P^{(-)} H P^{(-)}$ are two positive operators and $P^{(+)}$ ($P^{(-)}$) is the projector onto the subspace spanned by the eigenvectors of $H$ with positive (negative) eigenvalues. Note that $P^{(+)}P^{(-)}=0$ and hence $[P^{(+)},P^{(-)}]=0$ and $[ H^{(+)}, H^{(-)}]=0$.

Commutation of $H^{(+)}$ and $H^{(-)}$ implies that if we can exponentiate $H^{(+)}$ and $H^{(-)}$ separately, then we can also trivially realize $U_{\delta} = e^{-i\delta \,H}$ from concatenation of the unitary evolutions $U_{\delta}^{(\pm)}=e^{-i\delta \, H^{(\pm)}}$ as $U_{\delta}=U_{\delta}^{(-)} U^{(+)}_{\delta}$. To realize $U^{(+)}$ and $U^{(-)}$ we need to construct valid density matrices out of $H^{(\pm)}$as 
\begin{align}
\rho^{(\pm)}= H^{(\pm)}/\mathrm{Tr}[ H^{(+)}],
\end{align}
 such that 
\begin{align}
H=\mathrm{Tr}[ H^{(+)}]\rho^{(+)} - \mathrm{Tr}[ H^{(-)}]\rho^{(-)}.
\end{align} 
This brings $H$ in the form of Eq. \eqref{H-StateDecomposition}, where here the set $\{\rho_{j}\}=\{\rho^{(\pm)}\}$ depends on $H$.

\subsection{Shifting and normalizing a Hamiltonian}

Another possibility for writing any Hamiltonian $H$ in terms of states is to construct a quantum state (i.e., nonnegative, trace-unity operator) from the Hamiltonian as the following: First make the Hamiltonian positive by shifting its minimum eigenvalue $\lambda_{\min}$, i.e., $H \rightarrow H + \lambda_{\min} \openone $ and then making it a trace-unity operator by dividing it by its trace, $\rho_{H}$, i.e., $H + \lambda_{\min} \openone  \rightarrow \rho_{H} $, where 
\begin{align}
\rho_{H} = (H+\lambda_{\min} \openone )/(\mathrm{Tr}[H]+d\, \lambda_{\min}),
 \end{align}
 with $d$ being the dimension of the Hilbert space of the system. Thus, it is simple to see that, the Hamiltonian can be written in terms of $\rho_{H}$ as
\begin{align}
H = (\mathrm{Tr}[H]+d\, \lambda_{\min}) \rho_{H},
\label{rho_H}
\end{align}
modulo the dynamically trivial term $- \lambda_{\min} \openone$. Again, $H$ is now in the form of Eq. \eqref{H-StateDecomposition} and the single $\rho_{j}=\rho_{H}$ depends on $H$.
\subsection{Decomposition in terms of coherent states}

From the Glauber-Sudarshan $P$ representation, for any Hamiltonian $H$ one can find an $h(\alpha)$ function such that
\begin{align}
H= \textstyle{\int_{\mathds{C}}} d^{2}\alpha\, h(\alpha)\, |\alpha\rangle \langle \alpha|,
\end{align}
where $|\alpha\rangle$s are coherent states. Using the identities 
\begin{gather}
\langle \beta|\alpha \rangle  = e^{-(|\alpha|^{2}+|\beta|^{2} - 2\beta^{*}\alpha)/2},\\
\delta^{2}(\alpha)  = \delta(\mathrm{Re}[\alpha])\, \delta(\mathrm{Im}[\alpha]) =  (1/\pi^{2}) \textstyle{\int_{\mathds{C}}} d^{2} \beta \, e^{-\beta \alpha^{*} + \beta^{*}\alpha},
\end{gather}
where $\delta(x)$ is the Dirac delta function, and following the same steps for obtaining the $P$ representation of quantum states, we have
\begin{align}
h(\alpha)= \frac{e^{|\alpha|^{2}}}{\pi^{2}} \textstyle{\int_{\mathds{C}}} d^{2} \beta \,\langle -\beta| H |\beta\rangle\, e^{|\beta|^{2}-\beta \alpha^{*} + \beta^{*} \alpha} .
\end{align}

\textit{Remark.}---In the case the Hamiltonian $H=H(a, a^{\dag})$ is in the antinormal order $H=\sum_{mn} J_{mn}a^{m} a^{\dag n}$, by inserting the identity $\openone= (1/\pi)\int_{\mathds{C}} d^{2}\alpha\,|\alpha\rangle\langle \alpha|$ we obtain 
\begin{align}
H = (1/\pi) \textstyle{\sum_{mn}} J_{mn}\int_{\mathds{C}} d^{2}\alpha \, \alpha^{m} \alpha^{* n} |\alpha\rangle\langle \alpha|,
\end{align}
which is in the form of Eq. \eqref{H-StateDecomposition} with $\{\rho_{j}\}= \{|\alpha\rangle\langle \alpha|\}_{\alpha\in\mathds{C}}$.
\section{An alternative method for density matrix exponentiation}

First, another (auxiliary) quantum system with state $\rho$ is prepared. Next the exponential-\textsc{swap} gate (e-\textsc{swap}) $e^{-i\mathcal{S} \delta}$ is applied on the joint system state $\rho \otimes \sigma$. Here $\mathcal{S}$ is the \textsc{swap} gate and $\delta$ is a sufficiently small parameter. This yields for the state of the simulator 
\begin{align}
\hskip-3mm\mathrm{Tr}_{1}[e^{-i\mathcal{S} \delta} (\rho \otimes \sigma) e^{i\mathcal{S} \delta}] & = \sigma - i \delta [\rho, \sigma] + O(\delta^{2} \Vert \rho - \sigma \Vert^{2}) \nonumber\\
&= e^{- i \delta \rho} \sigma e^{i \delta \rho} + O(\delta^{2} \Vert \rho-\sigma \Vert^{2}),
\label{eq7}
\end{align} 
where $\mathrm{Tr}_{1}[\cdot]$ denotes partial tracing on the first (auxiliary) system, $\Vert \cdot\Vert$ is the standard operator norm, and we have used the \textsc{swap} identity $\mathrm{Tr}_{1}[\mathcal{S} (\rho\otimes \sigma)]=\rho \sigma$. Using $n$ copies of $\rho$ and repeating the above process $n$ times, where $h t= n \delta$, it is straightforward to show that 
\begin{align}
&\mathrm{Tr}_{1}\big[e^{-i\mathcal{S} \delta} \big(\rho \otimes \ldots \mathrm{Tr}_{1}[e^{-i\mathcal{S} \delta} (\rho \otimes \sigma) e^{i\mathcal{S}\delta}]\ldots\big) e^{i\mathcal{S} \delta}\big]\nonumber\\
&= e^{-i h t\,\rho}\,\sigma \, e^{i h t\,\rho}+ O(n \,\delta^{2} \Vert\rho-\sigma \Vert^{2}).
\label{eq:acc}
\end{align} 
To reach a desired simulation error $\epsilon$ (i.e., $n\delta^{2} \Vert\rho-\sigma \Vert^{2} \leqslant \epsilon$), the number of copies of auxiliary $\rho$ should be $n \geqslant O\big((\Delta t)^{2}/\epsilon\big)$.

\section{Details of Example 1: Parametric amplifier}
\label{app:NLC}

The Hamiltonian of an optical parametric amplifier can be rewritten in terms of coherent quantum states as \cite{IndCohwithoutIndEm}
\begin{equation}
H=\frac{1}{\pi^{2}} \textstyle{\int_{\mathds{C}}} \textstyle{\int_{\mathds{C}}} d^{2}\alpha \, d^{2}\beta\, \big(g\alpha^{*} \beta^{*} |\alpha\rangle_{s}\langle \alpha| \otimes |\beta\rangle_{i}\langle \beta|+ g^{*}\alpha\,\beta\,|\alpha\rangle_{s}\langle \alpha| \otimes |\beta\rangle_{i}\langle \beta| \big),
\label{eq:H}
\end{equation}
where $|\alpha\rangle_{s}$s and $|\beta\rangle_{i}$s are the coherent states with the signal and idler frequencies, respectively. We are interested in Hamiltonians where within the decomposition \eqref{eq:H} we have cutoff values $\alpha_{c}$ and $\beta_{c}$ such that only those coherent states for which $|\alpha|\leqslant |\alpha_{c}|$ and $|\beta|\leqslant |\beta_{c}|$ contribute to the integrals. By discretizing the integrals we obtain 
\begin{align}
H &\approx \frac{\Delta^{2}}{\pi^{2}}  \textstyle{\sum}_{k,j=1}^{N} \big(g\alpha_{k}^{*} \beta_{j}^{*} |\alpha_{k}\rangle_{s}\langle \alpha_{k}| \otimes |\beta_{j}\rangle_{i}\langle \beta_{j}|+ g^{*}\alpha_{k}\,\beta_{j}\,|\alpha_{k}\rangle_{s}\langle \alpha_{k}| \otimes |\beta_{j}\rangle_{i}\langle \beta_{j}| \big)+O(\Delta^{2}),\nonumber\\
\label{paramAmp}
&= \textstyle{\sum}_{k,j=1}^{N} h(\alpha_{k}, \beta_{j})+h^{\dag}(\alpha_{k}, \beta_{j})
\end{align}
where $N\Delta=\max\{|\alpha_{c}|,|\beta_{c}|\}$. Using the Trotter-Suzuki theorem it is evident that to simulate the above Hamiltonian we need to simulate an arbitrary term 
\begin{align}
h (\alpha,\beta) = (\Delta^{2}/\pi^{2})  g\alpha^{*} \beta^{*} |\alpha\rangle_{s}\langle \alpha| \otimes |\beta\rangle_{i}\langle \beta|.
\end{align} 

To do so we need an ancilla in the state $|\psi_{\delta}\rangle=|0\rangle -i(\Delta^{2}/\pi^{2}) g \alpha^{*}\beta^{*} \delta |1\rangle$ and put it together with four other systems such that the initial state of the total system is given by
\begin{align}
|\psi_{\delta}\rangle \otimes |\alpha\rangle_{s} \otimes |\beta\rangle_{i} \otimes |0\rangle_{s'} \otimes |0\rangle_{i'},
\end{align}
where $s'$ and $i'$, respectively, are two extra systems with the signal and idler properties forming the state of the simulator. Here we assume the initial state of the simulator is the vacuum state $|0\rangle$. By applying two c-\textsc{swap} gates with the ancilla as the control qubit and the signal systems as the target systems of the first c-\textsc{swap} gate and similarly the idler systems as the target systems of the second c-\textsc{swap} gate, we obtain 
\begin{align}
|0\rangle \otimes |\alpha\rangle_{s} \otimes |\beta\rangle_{i} \otimes |0\rangle_{s'} \otimes |0\rangle_{i'}
-i \frac{\Delta^{2}}{\pi^{2}}g \alpha^{*}\beta^{*} \delta |1 \rangle \otimes \mathcal{S}_{ii'} \mathcal{S}_{ss'} \big(|\alpha\rangle_{s} \otimes |\beta\rangle_{i} \otimes |0\rangle_{s'} \otimes |0\rangle_{i'} \big).
\end{align}
Now we apply a selective measurement on the $\{|\pm\rangle\}$ basis and keep the state only when the result is $|+\rangle$, which (up to normalization) yields 
\begin{align}
&|+\rangle \otimes \Big(|\alpha\rangle_{s} \otimes |\beta\rangle_{i} \otimes |0\rangle_{s'} \otimes |0\rangle_{i'} -i \frac{\Delta^{2}}{\pi^{2}}g \alpha^{*}\beta^{*} \delta |0\rangle_{s} \otimes |0\rangle_{i} \otimes |\alpha\rangle_{s'} \otimes |\beta\rangle_{i'} \Big).
\end{align}
Finally, by discarding the ancilla, signal $s$, and idler $i$ systems (i.e., by tracing over them) we arrive at the desired simulated dynamics (up to $O(\delta^{2})$),
\begin{align}
\label{NewSimState-1}
 |0 \rangle_{s'}\langle 0| &\otimes |0 \rangle_{i'}\langle 0| -i \frac{\Delta^{2}}{\pi^{2}}g \alpha^{*}\beta^{*} \delta 
|\alpha \rangle_{s'}\langle \alpha |0 \rangle_{s'}\langle 0| \otimes |\beta\rangle_{i'}\langle \beta |0 \rangle_{i'}\langle 0|
+i \frac{\Delta^{2}}{\pi^{2}}g \alpha \beta \delta |0 \rangle_{s'}\langle 0|\alpha\rangle_{s'}\langle \alpha| \otimes |0 \rangle_{i'}\langle 0|\beta\rangle_{i'}\langle \beta|a.
\end{align}
From Eq. \eqref{NewSimState-1} and using the identity $\langle 0|\alpha\rangle=e^{-|\alpha|^2/2}$ and the similar one for $\langle 0|\beta\rangle$, it is immediate to show that the state of the simulator after applying the simulation protocol for all terms in the Hamiltonian becomes
\begin{align}
\label{NewSimState-2}
\big( |0 \rangle_{s'} |0 \rangle_{i'}-i \frac{\Delta^{2}}{\pi^{2}}g \delta \textstyle{\sum_{k, j}} \alpha_k^{*}\beta_j^{*} \delta e^{-(|\alpha_k|^2+|\beta_j|^2)/2} |\alpha\rangle_{s'}|\beta\rangle_{i'}\big)\big(h.c.\big)+O(\delta^{2}) \approx \big( |0 \rangle_{s'} |0 \rangle_{i'}-i g \delta |1 \rangle_{s'} |1 \rangle_{i'}\big)\big(h.c.\big)+O(\delta^{2}),
\end{align}
where we have used $|1\rangle= \frac{1}{\pi} \int_0^{\infty}  |\alpha\rangle \langle\alpha |1\rangle \approx \frac{\Delta}{\pi}\textstyle{\sum_{k}} \alpha_k^{*} e^{-|\alpha_k|^2/2} |\alpha_k\rangle$, where we have discretized the integral and used the identity $\langle\alpha |1\rangle =\alpha^{*} e^{-|\alpha|^2/2}$. Equation \eqref{NewSimState-2} means that a pair of signal-idler photons are created. This is exactly what we expect from a parametric amplifier when applied on a vacuum state. 

It should be noted that Eq. \eqref{NewSimState-1} can also be written approximately as 
\begin{align}
e^{-i \frac{\Delta^{2}}{\pi^{2}} g \alpha^{*}\beta^{*} \delta |\alpha\rangle_{s'}\langle \alpha| \otimes |\beta \rangle_{i'}\langle \beta|}
\big(|0 \rangle_{s'}\langle 0|\otimes |0 \rangle_{i'}\langle 0|\big) e^{i \frac{\Delta^{2}}{\pi^{2}} g \alpha \beta \delta |\alpha\rangle_{s'}\langle \alpha| \otimes |\beta \rangle_{i'}\langle \beta|} + O(\delta^{2}).
\end{align}
By repeating similar procedures for simulating other terms in Eq. \eqref{paramAmp} on the new state of the simulator in Eq. \eqref{NewSimState-2} we can simulate the whole Hamiltonian of the parametric amplifier. 

\section{Example: A two-level atom with open-system evolution}
\label{app:opsys}

Assume an open quantum system evolving as $\dot{\sigma}=-i[\omega_{0} S_{z},\sigma]+\gamma (S_{x} \sigma S_{x}-\sigma)$, where $S_{x}$, $S_{y}$, and $S_{z}$ are the Pauli matrices. Thus, the vectorized Lindbladian is given by $\mathbbmss{L}=-i\omega_{0} (S_{z} \otimes \openone- \openone \otimes S_{z})+\gamma (S_{x} \otimes S_{x}-\openone\otimes \openone)$. Since $S_{z}=2 |0\rangle\langle 0|-\openone$ and $S_{x}=2 |+\rangle\langle +|-\openone$, where $ |+\rangle=( |0\rangle+ |1\rangle)/\sqrt{2}$, $S_{z} |0\rangle=|0\rangle$ and $S_{z} |1\rangle=-|1\rangle$, we can obtain the decomposition in terms of quantum states as  $\mathbbmss{L} = -2i\omega_{0} (|0\rangle\langle 0| \otimes \openone- \openone \otimes |0\rangle\langle 0|)+ 4\gamma |+\rangle\langle +|\otimes |+\rangle\langle +|- 2\gamma |+\rangle\langle +| \otimes \openone-2\gamma \openone \otimes |+\rangle\langle +|$. 

Using the identities $S_{z}^{T}=S_{z}$ and $S_{x}^{T} = S_{x}$ in the computational basis, we can obtain the vectorized Lindblad equation as
\begin{equation}
|\dot{\sigma}\rrangle=-i\omega_{0} (S_{z} \otimes \openone- \openone \otimes S_{z})|\sigma\rrangle + \gamma (S_{x} \otimes S_{x}-\openone\otimes \openone)|\sigma\rrangle. 
\label{nu-dyn}
\end{equation}
To decompose the vectorized Lindbladian $\mathbbmss{L}$ in terms of quantum states we use the identities $S_{z}=2 |0\rangle\langle 0|-\openone$ and $S_{x} = 2 |+\rangle\langle +|-\openone$, where $ |+\rangle = (|0\rangle + |1\rangle)/\sqrt{2}$, $S_{z} |0\rangle=|0\rangle$, and $S_{z} |1\rangle = -|1\rangle$. Thus, $\mathbbmss{L}$ is decomposed in terms of quantum states as 
\begin{equation}
 \mathbbmss{L}=-2i\omega_{0} \big(|0\rangle\langle 0| \otimes \openone- \openone \otimes |0\rangle\langle 0|)+ 4\gamma |+\rangle\langle +|\otimes |+\rangle\langle +|- 2\gamma |+\rangle\langle +| \otimes \openone-2\gamma \openone \otimes |+\rangle\langle +| + q \openone \otimes\openone, 
\end{equation}
where the last term is proportional to the identity with some coefficient $q$ which depends on $\omega_{0}$ and $\gamma$. Since this term does not contribute to the dynamics, it is irrelevant for the simulation and we can drop it from $ \mathbbmss{L}$, whence
\begin{equation}
\mathbbmss{L}=-2i\omega_{0} \big(|0\rangle\langle 0| \otimes \openone- \openone \otimes |0\rangle\langle 0|\big)+ 4\gamma |+\rangle\langle +|\otimes |+\rangle \langle +| - 2\gamma  |+\rangle\langle +| \otimes \openone-2\gamma \openone \otimes |+ \rangle \langle +|.
\end{equation}
 
 Assuming that the system is initially in the pure state $|+\rangle$, its counterpart in the vectorized space becomes 
\begin{align}
|\sigma(0)\rrangle=\frac{1}{2}\left( \begin{array}{c} 1\\ 1\\ 1\\ 1 \end{array} \right).
\end{align}
Note that a pure state remains normalized when vectorized (due to $\sigma^{2}=\sigma$). According to the nonunitary dynamics (\ref{nu-dyn}) the state evolves as 
\begin{align}
|\sigma(t)\rrangle = \frac{1}{2}\left(\begin{array}{c} 1 \\ e^{-\gamma t} \big(\cosh(\Omega t) + (1/\Omega)(\gamma -2 i \omega_{0}) \sinh(\Omega t) \big) \\  e^{-\gamma t} \big(\cosh(\Omega t) + (1/\Omega)(\gamma +2 i \omega_{0}) \sinh(\Omega t) \big) \\ 1 \end{array} \right), \end{align}
where $\Omega = \sqrt{\gamma^{2} - 4 \omega_{0}^{2}}$ and the state is not normalized. Because
\begin{equation}
\llangle \sigma(t)|\sigma(t)\rrangle=\frac{\gamma  e^{-2 \gamma t} \big(\gamma  \cosh(2\Omega t) + \Omega \sinh(2\Omega t) \big) + \gamma^{2} -4 \omega_{0}^{2} \left(e^{-2\gamma t} + 1 \right)}{2\Omega^{2}},
\end{equation}
the normalized state based on which the results of the measurement in the lab is obtained is given by $|\psi(t)\rangle= |\sigma(t) \rrangle/\sqrt{\llangle \sigma(t)|\sigma(t)\rrangle}$. From $(\langle 0| \otimes \langle 0|)|\sigma(t)\rrangle+(\langle 1|\otimes \langle 1|)|\sigma(t)\rrangle=1$ the state of the original system is obtained as 
\begin{align}
\sigma(t) = \frac{1}{2}\left( \begin{array}{cc} 1 & e^{-\gamma t} \left(\cosh(\Omega t) + \frac{(\gamma -2 i \omega_{0})}{\Omega} \sinh(\Omega t) \right) \\ e^{-\gamma t} \left(\cosh (\Omega t)+\frac{(\gamma +2 i \omega_{0})}{\Omega} \sinh(\Omega t)\right) & 1 \end{array} \right).
\end{align}

\section{Details of the calculations of the updated control system}
\label{app:updated-control}

We have 
\begin{align}
\xi \otimes \DSGamma & = \xi \otimes\underset{n_{1}}{\underbrace{\sigma(\tau-a_{1})\otimes\cdots}}\otimes \cdots \otimes \underset{n_{i}}{\underbrace{\sigma(\tau-a_{i})\otimes\cdots}}\otimes \cdots
\end{align}
Thus, we have $N+1$ target systems labelled by $0$ to $N$. We then apply $N$ separate c-\textsc{swap} gates $U_{\text{cs}_{i,j}}$, where $i$ and $j$ indicate on which target systems \textsc{swap} is applied,
\begin{align}
U_{\text{cs}_{N-1,N}}\cdots&\,\, U_{\text{cs}_{1,2}} U_{\text{cs}_{0,1}} \big(|\psi_{\delta}\rangle\langle \psi_{\delta}|\otimes \xi \otimes\sigma(\tau-a_{1}) \otimes\cdots \otimes \sigma(\tau-a_{2})\otimes \cdots \otimes 
\sigma(\tau-a_{i})\otimes\cdots
\otimes \cdots\big) U^{\dag}_{\text{cs}_{0,1}} U^{\dag}_{\text{cs}_{1,2}} \cdots U^{\dag}_{\textsc{cs}_{N-1,N}}
\nonumber\\
& = |0\rangle\langle 0| \otimes \xi \otimes \sigma(\tau-a_{1})\otimes\cdots\otimes 
\sigma(\tau-a_{2})\otimes\cdots \otimes \sigma(\tau-a_{i})\otimes\cdots \nonumber\\
& \,\,\,\,\, -ic\delta |1\rangle\langle 0|\otimes \mathcal{S}_{N-1,N} \cdots \mathcal{S}_{1,2} \mathcal{S}_{0,1} \xi \otimes \sigma(\tau-a_{1})\otimes\cdots\otimes \sigma(\tau-a_{2})\otimes\cdots \otimes \sigma(\tau-a_{i})\otimes\cdots \nonumber\\
& \,\,\,\,\, + ic\delta |0\rangle\langle 1|\otimes \xi \otimes \sigma(\tau-a_{1})\otimes\cdots\otimes \sigma(\tau-a_{2})\otimes\cdots \otimes  \sigma(\tau-a_{i}) \otimes \cdots \mathcal{S}_{0,1} \mathcal{S}_{1,2}\cdots \mathcal{S}_{N-1,N} + O(\delta^{2}).
\end{align}
We then trace over all target systems and obtain the updated control system as 
\begin{align}
|\tilde{\psi}\rangle\langle \tilde{\psi}|=& \, \mathrm{Tr}_{\textsc{target}}\big[|0\rangle\langle 0| \otimes \xi \otimes \sigma(\tau-a_{1})\otimes\cdots \otimes \sigma(\tau-a_{2})\otimes \cdots \otimes  \sigma(\tau-a_{i})\otimes\cdots \nonumber\\
&\, -ic\delta |1\rangle\langle 0|\otimes \mathcal{S}_{N-1,N}\cdots \mathcal{S}_{1,2} \mathcal{S}_{0,1} \xi \otimes \sigma(\tau-a_{1}) \otimes \cdots \otimes \sigma(\tau-a_{2})\otimes\cdots \otimes \sigma(\tau-a_{i})\otimes\cdots \nonumber\\
&\, +ic\delta |0\rangle\langle 1|\otimes \xi \otimes \sigma(\tau-a_{1})\otimes\cdots \otimes \sigma(\tau-a_{2})\otimes\cdots \otimes \sigma(\tau-a_{i}) \otimes \cdots \mathcal{S}_{0,1} \mathcal{S}_{1,2}\cdots \mathcal{S}_{N-1,N} 
\big] + O(\delta^{2}) \nonumber\\
= &\, |0\rangle\langle 0| -ic\delta |1\rangle\langle 0| \, \mathrm{Tr}\big[\xi \sigma^{n_{1}}(\tau-a_{1})\sigma^{n_{2}}(\tau-a_{2})\cdots \sigma^{n_{i}}(\tau-a_{i})\cdots\big] \nonumber\\
&\, +ic\delta |0\rangle\langle 1| \, \mathrm{Tr}\big[\cdots \sigma^{n_{i}}(\tau-a_{i})\cdots \sigma^{n_{2}}(\tau-a_{2})\sigma^{n_{1}}(\tau-a_{1}) \xi\big] +O(\delta^{2}),
\end{align}
which is equivalent to 
\begin{align}
|\tilde{\psi}\rangle\langle \tilde{\psi}|=|0\rangle\langle 0| -ic \delta |1\rangle\langle 0| \, \mathrm{Tr}[\xi \Gamma]+ic\delta |0\rangle\langle 1| \, \mathrm{Tr}[\xi \Gamma]^{*} + O(\delta^{2}),
\end{align}
or
\begin{equation}
|\tilde{\psi}\rangle = |0\rangle + ic\delta\,\mathrm{Tr}[\xi \Gamma]|1\rangle. 
\end{equation}

\section{Details of Example 2: Gross-Pitaevskii Hamiltonian}
\label{app:GP}

The Gross-Pitaevskii equation for the wave function $\psi(r,t)$ of a particle of mass $m$ under the position (or space) $r$-dependent potential $V(r)$ is given by 
\begin{align}
i\dot{\psi}(r,t)= \big(-\nabla^{2}/2m+V(r)+g |\psi(r,t)|^{2}\big) \,\psi(r,t).
\end{align}
By rewriting the operator form of the above elementwise Hamiltonian we obtain that 
\begin{align}
H=&\,\,  H_{0} + g \textstyle{\sum_{r}} |\psi(r,t)|^{2} |r\rangle\langle r|\nonumber\\
=& \,\, H_{0} + g \textstyle{\sum_{r}} |\langle r|\psi(t)\rangle|^{2} |r\rangle\langle r|\nonumber\\
=& \,\, H_{0} + g \textstyle{\sum_{r}} |r\rangle\langle r|\psi(t)\rangle\langle \psi(t)|r\rangle\langle r|\nonumber\\
=& \,\,  H_{0} + g \textstyle{\sum_{r}} \Pi_{r} |\psi(t)\rangle\langle \psi(t)| \Pi_{r},
\end{align}
where for simplicity we have assumed that the space is discrete. In the above equation $H_{0} = -\nabla^{2}/(2m)+\sum_{r}  V(r) |r\rangle\langle r|$ is the state-independent (linear) part of the Hamiltonian and the projectors $\Pi_{r} $ are given by $\Pi_{r} = |r \rangle\langle r|$.

\section{SBQS of nonlinear delay-differential equations}

We propose a SBQS algorithm to simulate NLDs we need to follow several steps:\\

(\textbf{i}) We add two extra equations to the set in Eq. \eqref{NLDDEq}. One equation is added for a new independent fixed parameter $x_{0}$: $\dot{x}_{0}=0$, which implies $f_{00}=1$ and $f_{m0}=f_{0n}=0,\,\forall m,n$. This constant parameter has been introduced to take care of some technicalities which will become clear later. Another equation is added for a normalizing variable $x_{D+1}=\textstyle{\sqrt{1-\sum_{i=0}^{D} x_{i}^{2}}}$, such that $\dot{x}_{D+1} = -x_{D+1}^{-1} \boldsymbol{x} \cdot \dot{\boldsymbol{x}} = -x_{D+1}^{-1} \sum_{m,n=1}^{D} f_{mn} x_{m} x_{n}=:f_{D+1,0}\,x_{0}$, where $f_{D+1,0} =-(x_{0} x_{D+1})^{-1} \sum_{m,n=1}^{D} f_{mn} x_{m} x_{n}$. Using the Taylor expansion about $x_{D+1}=1$ we obtain $x_{D+1}^{-1}= \sum_{k=0}^{r} (1-x_{D+1})^{k} + O\big((1-x_{D+1})^{r}\big)$, where $r$ is chosen according to our desired error. Thus,  
$f_{D+1,0}$ becomes a polynomial of all variables similar to the expression for $f_{mn}$ in Eq. \eqref{f_{m}n-poly}, but where index $i$ runs from $0$ to $D+1$. If necessary, we can first scale all variables so that normalization becomes possible. Now we define a normalized quantum state as
\begin{align}
\label{var_vec}
|\psi(t)\rangle=\big(x_{0}, x_{1}(t),\cdots,x_{D}(t), x_{D+1}(t)\big)^{T} = \textstyle{\sum_{i=0}^{D+1}} x_{i}(t)|i\rangle.
\end{align}
It is straightforward to see that the set of differential equations can be recast in the form of a Schr\"{o}dinger-like equation $|\dot{\psi}(t)\rangle=-i H_{\textsc{nld}} |\psi(t)\rangle$, where  
\begin{align}
H_{\textsc{nld}}=i\textstyle{\sum}_{m,n=0}^{D+1} f_{mn}\big(\boldsymbol{x}(t),\boldsymbol{x}(t-a_{1}),\cdots, \boldsymbol{x}(t-a_{N})\big) |m\rangle\langle n|
\label{H-nonlinearEq}
\end{align}
is a \textit{nonlinear} (variable-dependent) operator which is not necessarily Hermitian. Nonetheless, SBQS can successfully simulate its dynamics.\\ 

(\textbf{ii}) To transform $H_{\textsc{nld}}$ into the form of Eq. \eqref{H-StateDecomposition} with nonlinear couplings, it is sufficient to write $f_{mn}$ in the form of Eq. \eqref{hNL}. We note that if $r_{ij}$ is even, then $x_{i}^{r_{ij}}(t-a_{j})=\big(\langle \psi(t-a_{j})|i \rangle\langle i|\psi(t-a_{j}) \rangle\big)^{r_{ij}/2}$, which is the expectation value of the operator $|i\rangle\langle i|^{\otimes r_{ij}/2}$ with respect to multiple copies of the state of the system at relevant times, $|\psi(t-a_{j})\rangle^{\otimes r_{ij}/2}$. When $r_{ij}$ is odd, using the constant parameter $x_{0}$ we can also write $x_{i}^{r_{ij}}(t-a_{j})$ in terms of the expectation value of an operator: $x_{i}^{r_{ij}}(t-a_{j})= \big(\langle \psi(t-a_{j})|i \rangle\langle i|\psi(t-a_{j})\rangle\big)^{\lfloor {r_{ij}}/{2} \rfloor} \big(x_{0}^{-1}\langle \psi(t-a_{j})|0 \rangle\langle i|\psi(t-a_{j})\rangle\big)$, which is the expectation value of the non-Hermitian operator 
$x_{0}^{-1}|i\rangle\langle i|^{ \otimes\lfloor r_{ij}/2\rfloor}\otimes |0\rangle\langle i|$ with respect to the state $|\psi(t-a_{j}) \rangle^{\otimes\lfloor r_{ij}/2\rfloor+1}$. Here $\lfloor \cdot \rfloor$ denotes the floor function. For simplicity and without loss of generality, we assume in the following that all $r_{ij}$s are even. Hence we can rewrite any $f_{mn}$ given in Eq. \eqref{f_{m}n-poly} as the expectation value of the operator 
\begin{align}
F_{mn}=\textstyle{\sum}_{\{r_{ij}\}=\{1\}}^{\{\ell_{j}\}} C^{mn}_{\{r_{ij}\}} \textstyle{\otimes_{j=0}^{N} \otimes_{i=0}^{D+1} }  | i\rangle\langle i|^{\otimes r_{ij}/2},
\end{align}
with respect to the quantum state
\begin{align}
\label{Psi}
|\Psi\rangle:=\textstyle{\otimes}_{j=0}^{N} |\psi(t-a_{j}) \rangle^{\otimes  (D+2) \ell_{j}/2},
\end{align}
such that $f_{mn}=\langle\Psi|F_{mn}|\Psi \rangle$. Equivalently, using the \textsc{swap} identity we can rewrite 
\begin{align}
f_{mn}=\mathrm{Tr}[\mathcal{S}\,(F_{mn}\otimes|\Psi \rangle\langle\Psi|)].
\end{align}
Using this, it is evident that all terms in $H_{\textsc{nld}}$ can be written in the desired form of Eq. \eqref{NLH} and hence can be simulated using SBQS.\\ 

As a final remark in this part, note that because in the steps of the simulation explained in Sec. \ref{sec:nonlinear} all traces are taken over different parties, we can alternatively defer all traces until the end of the protocol. In addition, SBQS bypasses the need for computation of intermediate averages by a suitable post-selection [Eq. (\ref{post-selection})]. These features may offer some advantages for practical implementations of the simulation---see, e.g., remarks made in Ref. \cite{Molmer-NME2}.

\ignore{
\textit{Example 4: Logistic map}. Consider the logistic map differential equation $\dot{x}= (r-1) x-r x^{2}$. 
Here we need three copies of the state of the system at each time to simulate every single term of the Trotter-Suzuki expansion, which includes several terms. It should be noted that the number of Trotter-Suzuki terms depends on the decomposition of the Hamiltonian in terms of quantum states. Here we have performed this with a brute-force approach and as a proof of principle in order to demonstrate the direct, explicit implementation of SBQS. Details have been relegated to SI. 
}

\section{Details of Example 3: Logistic map}
\label{app:LM}

Here we explain the details of applying the SBQS method for the logistic map (LM) differential equation $\dot{x}= (r-1) x-r x^{2}$. We first scale the variable so that it becomes normalizable for a reasonable range, i.e., $x \to \alpha x$,
\begin{align}
\dot{x}= (r-1) x-r \alpha x^{2} .
\label{logistic}
\end{align}
This differential equation does not have any time history and is local in time. Hence we need copies of the system only at $t$.
Now we define the quantum state 
\begin{align}
|\psi_{\textsc{lm}}(t)\rangle=\left(\begin{matrix}
 x_{0} \\
 x_{1}(t) \\
 x_{2}(t)
\end{matrix} \right), 
\end{align}
such that $x_{1}(t)=x(t)$, $x_{2}=\sqrt{1-x_{0}^{2}- x^{2}}$. To obtain $\dot{x}_{2}$ we use the Taylor expansion $1/x_{2}=3 -3x_{2} +x_{2}^{2} + O\big((1-x_{2})^{3}\big)$ and obtain $\dot{x}_{2} \approx -(3 -3x_{2} +x_{2}^{2}) \big( (r-1) x^{2}-r \alpha x^{3}\big)$. Thus, 
\begin{align}
H_{\textsc{lm}} = \left(\begin{matrix} 0 & 0 & 0 \\ 0 & r-1-r \alpha x  & 0 \\ -x_{0}^{-1} (3 -3\,x_{2} +x_{2}^{2}) \big((r-1) x^{2}-r \alpha x^{3}\big) & 0 & 0 \end{matrix}\right), \end{align}
from which
\begin{align}
f_{11}=&\, (r-1) -r \alpha x, \\
f_{20}=&\, -3 x_{0}^{-1} (r-1) x^{2} + 3 r x_{0}^{-1} \alpha x^{3} +3 x_{0}^{-1}  (r-1)  x^{2} x_{2} - 3r \alpha x_{0}^{-1}  x^{3} x_{2}-x_{0}^{-1} (r-1) x^{2} x_{2}^{2}+ r \alpha x_{0}^{-1} x^{3} x_{2}^{2}.
\end{align}
These relations lead to
\begin{align}
\label{g6}
F_{11} = &\, (r-1) x_{0}^{-2}|0\rangle\langle 0|\otimes \openone\otimes \openone -r \alpha x_{0}^{-1} |0\rangle\langle 1|\otimes \openone\otimes \openone, \\
\label{g7}
F_{20} = &\, -3 x_{0}^{-1} (r-1) |1\rangle\langle 1| \otimes \openone \otimes \openone+ 3 r x_{0}^{-2} \alpha |1\rangle\langle 1|\otimes |0\rangle\langle 1| \otimes \openone+3 x_{0}^{-2}  (r-1) |1\rangle\langle 1|\otimes |0\rangle\langle 2|\otimes \openone \nonumber\\
&\, - 3r \alpha x_{0}^{-3} |1\rangle\langle 1|\otimes |0\rangle\langle 1|\otimes |0\rangle\langle 2|-x_{0}^{-1} (r-1) |1\rangle\langle 1|\otimes |2\rangle\langle 2|\otimes \openone + r \alpha x_{0}^{-2} |1\rangle\langle 1|\otimes |0\rangle\langle 1|\otimes |2\rangle \langle 2|,
\end{align}
while the rest of $F_{mn}$s vanish. 
Thus, to simulate this evolution we need three copies of the state of the system at each time, which means
\begin{align}
|\Psi_{\textsc{lm}} (t) \rangle = |\psi_{\textsc{lm}}(t)\rangle \otimes |\psi_{\textsc{lm}}(t)\rangle \otimes |\psi_{\textsc{lm}}(t)\rangle.
\end{align}
According to the definitions \eqref{polarization_bases} we can expand $|0\rangle\langle 1|$ and $|0\rangle\langle 2|$ in terms of relevant quantum states,
\begin{align}
|0\rangle\langle 1|=&-\textstyle{\frac{1+i}{2}} |0 \rangle \langle 0| -\frac{1+i}{2} |1\rangle\langle 1| + |+_{01} \rangle \langle +_{01}| + i |-_{01}\rangle  \langle -_{01} |,\nonumber\\
|0\rangle\langle 2|=& -\textstyle{\frac{1+i}{2}} |0 \rangle \langle 0| -\frac{1+i}{2} |2\rangle \langle 2| + |+_{02} \rangle \langle +_{02}| + i |-_{02} \rangle \langle -_{02}|.
\end{align}
Inserting these equations into Eqs. \eqref{g6} and \eqref{g7} we see that $F_{11}$ is expanded in terms of five tripartite quantum states and $F_{20}$ is expanded in terms of thirty tripartite quantum states. For example, we obtain
\begin{align}
F_{11} = & \, (r-1) x_{0}^{-2}|0 \rangle \langle 0|\otimes \openone\otimes \openone +r \alpha x_{0}^{-1} \textstyle{\frac{1+i}{2}} |0\rangle \langle 0| \otimes \openone\otimes \openone + r \alpha x_{0}^{-1} \frac{1+i}{2} |1 \rangle \langle 1| \otimes \openone \otimes \openone-r \alpha x_{0}^{-1}|+_{01} \rangle \langle +_{01}| \otimes \openone \otimes \openone \nonumber\\
&- i r \alpha x_{0}^{-1} |-_{01} \rangle \langle -_{01}|\otimes \openone \otimes \openone,
\end{align}
and hence its expectation value with respect to $|\Psi(t)\rangle$ leads to the desired form
\begin{align}
f_{11}=&\, (r-1) x_{0}^{-2} \,\mathrm{Tr} \big[\mathcal{S} \left( |0\rangle\langle 0|\otimes |\psi_{\textsc{lm}}(t)\rangle \langle\psi_{\textsc{lm}}(t)| \right) \big]+ r \alpha x_{0}^{-1} \textstyle{\frac{1+i}{2}} \,\mathrm{Tr} \big[ \mathcal{S} \left(|0 \rangle \langle 0| \otimes |\psi_{\textsc{lm}}(t) \rangle \langle \psi_{\textsc{lm}}(t)| \right) \big] \nonumber\\
&\, + r \alpha x_{0}^{-1} \textstyle{\frac{1+i}{2}} \,\mathrm{Tr} \big[ \mathcal{S} \left(|1 \rangle \langle 1| \otimes |\psi_{\textsc{lm}}(t) \rangle \langle \psi_{\textsc{lm}}(t)| \right)\big] - r \alpha x_{0}^{-1} \,\mathrm{Tr} \big[ \mathcal{S} \left(|+_{01}\rangle \langle +_{01}|\otimes |\psi_{\textsc{lm}} (t) \rangle\langle\psi_{\textsc{lm}}(t) |\right) \big] \nonumber\\
&\, - i r \alpha x_{0}^{-1} \,\mathrm{Tr} \big[ \mathcal{S} \left(|-_{01}\rangle\langle -_{01}|\otimes |\psi_{\textsc{lm}}(t) \rangle \langle \psi_{\textsc{lm}}(t)|\right) \big].
\end{align}
Since the Hamiltonian is $H_{\textsc{lm}} = f_{11}|1\rangle \langle 1| - f_{20}|2\rangle \langle 0|$, we need to also expand $|2\rangle \langle 0|$ in terms of  quantum states. Hence overall the Trotter-Suzuki expansion leads to $5+ 30 \times 4$ terms, each of which needs to be simulated by SBQS. Considering the number of copies needed for each time simulation, it can be calculated that to simulate the dynamics up to time $t=n \delta$ the number of simulation steps is of the order of $O(n R C^{n})$, where $C$ is the number of needed copies, here $C=3$, and $R$ is the number of simulation steps for the given $C$, here $R= 5+30 \times 4$. It is evident from Eq. \eqref{Psi} in the main text that, generally the number of copies is $C= O\big(N(D+2)\ell/2\big)$, where $\ell = \max_{j=1}^{N} \{\ell_{j}\}$ and it increases in time  exponentially as $C(n) = O\big([N(D+2)\ell/2]^{n}\big)$. This exponential behavior is inevitable for general nonlinear differential equations  \cite{Childs-1,Childs-dissipative-NL}. 

\section{Proof that the extended state-dependent CD evolution for pure states preserves the eigenstates}
\label{app:proof-cd}

A \textit{pure} quantum state which satisfies the CD dynamical equation $\dot{\sigma}(t)=-i[H(t), \sigma(t)]+\big[[\dot{\sigma}(t), \sigma(t)],\sigma(t)\big]$ yields $[H(t), \sigma(t)]=0$. This can be seen using the relation $\sigma^{2}=\sigma$, hence $\dot{\sigma}\sigma+\sigma \dot{\sigma}=\dot{\sigma}$ or equivalently $\sigma \dot{\sigma} \sigma=0$ and
\begin{equation}
\big[[\dot{\sigma}, \sigma],\sigma\big]= \dot{\sigma}\sigma-\sigma \dot{\sigma}\sigma-\sigma\dot{\sigma}\sigma +\sigma \dot{\sigma} = \dot{\sigma},\,\,\forall t.
\end{equation}
Hence $[H(t),\sigma(t)]=0$. Since we have assumed that $\sigma(t)$ is a pure state, we can write it as $\sigma(t) = |\psi(t)\rangle\langle \psi(t)|$ for some $|\psi(t)\rangle$, and then we obtain the equation
\begin{align}
H(t)|\psi (t)\rangle=\langle \psi (t) | H(t)| \psi(t) \rangle \, |\psi(t)\rangle.
\label{g1}
\end{align}
This equation means that the instantaneous state $|\psi(t)\rangle$ of the system should always be an eigenstate of the instantaneous Hamiltonian $H(t)$. This result remains unchanged whether the Hamiltonian $H(t)$ is state dependent or not. 

If $H(t)$ has a nondegenerate eigenstate at all times, then obviously Eq. \eqref{g1} implies that if initially the system is, e.g., in the ground state, the state of the system under this evolution remains always in the instantaneous ground state. If we are interested only in the ground state, then only the condition of nonvanishing gap with the excited eigenspace suffices and we are guaranteed to remain in the ground eigenspace. This result can also be extended in a straightforward manner to the case of degenerate Hamiltonians \cite{Refael-etal}. 

\section{Adiabatic evolution for optimization problems: Variance}
\label{app:variance}

Assume we want to minimize the variance $(\Delta A)^{2} = \langle A^{2}\rangle - \langle A \rangle^{2}$ of a given observable $A$, where $\langle A\rangle = \mathrm{Tr}[A\Xi]$ and $\Xi$ is the state of the system. The goal then is to obtain $\min_{\Xi}(\Delta A)^{2}$. The solution is any eigenstate of the observable $A$, which are assumed unknown or difficult to obtain.

We note that we can rewrite $(\Delta A)^{2} = \mathrm{Tr}\big[\big((A^{2}\otimes \openone + \openone \otimes  A^{2})/2 - A \otimes A\big) \Xi \otimes \Xi \big]$. This implies that the optimized state is the ground state of $H_{A}=(A^{2} \otimes \openone + \openone \otimes A^{2})/2 - A \otimes A$. Since the ground state is not necessarily in the form of a separable state, using it yields a \textit{lower bound} on $\min_{\Xi}(\Delta A)^{2}$. One approach to obtain the ground state is to transform this problem into a quantum adiabatic evolution or state preparation. To do so, consider a quantum system with some initial state $\sigma(0) = |\phi_{0}\rangle \langle \phi_{0}|$, where $|\phi_{0}\rangle$ is the ground state of an ``easy'' Hamiltonian $H_{0}$. Next prepare the interpolating Hamiltonian $H\big(s(t)\big) = \big(1-s(t)\big) H_{0}+s(t)\, H_{A}$, with $T$ a sufficiently large time at which $\sigma(T)\approx \arg\min_{\Xi}(\Delta A)^{2}$, and $s(t)$ is a sufficiently smooth knob function satisfying the boundary conditions $s(0)=1-s(T)=0$, while also keeping the energy gap of $H\big(s(t)\big)$ always open \cite{exp-ad}. Instead, we can solve this problem by using SBQS. Following Eq. (\ref{CD-Ham}), we can simulate the counterdiabatic \textit{state-dependent} Hamiltonian $\widetilde{H}_{\textsc{cd}}(t,\sigma) = H\big(s(t)\big) - (i/\tau)[\sigma(t-\tau), \sigma(t)]$ with arbitrary $T>0$. This directly yields a lower bound on $\min_{\Xi}(\Delta A)^{2}$.

Additionally, rather than lower bounds, we can also obtain the \textit{exact} optimized state. First rewrite the variance as $(\Delta A)^{2} = \mathrm{Tr}[(A^{2}-\langle A \rangle A) \Xi]$. This implies that the optimal state is the ground state of the \textit{state-dependent} Hamiltonian $H'_{A}=A^{2}+\langle A\rangle A$. To attain this state, one can follow the steps described above for the state-independent Hamiltonian $H_{A}$: start with an easy Hamiltonian $H_{0}$ and its ground state $|\phi_{0}\rangle$; next simulate the counterdiabatic evolution related to the transient Hamiltonian $H\big(s(t)\big)=\big(1-s(t)\big)H_{0}+s(t) \,H'_{A}$. However, here due to the state dependence the choice of the initial Hamiltonian $H_{0}$ (and its ground state) affects the final Hamiltonian and its ground-state energy. To remove this dependence, we also need to perform an optimization over all initial Hamiltonians.

\textit{Remark}.---Here we have presented a number of ideas and methods specially adapted for use in SBQS. But we envision that some of these specific proposals may also find applications in other quantum and classical approaches to solve similar problems.

\end{document}